\documentclass[hyper]{JHEP3}

\input{epsf}
\usepackage{epsfig}
\usepackage{amssymb}
\usepackage{amsmath}
\usepackage{amsfonts}
\usepackage{textgreek,upgreek}
\usepackage{amsbsy}
\usepackage[all]{xy}
\usepackage{amsmath}

\usepackage{amssymb,amscd}
\usepackage{mathrsfs}
\usepackage{amsmath,amsthm}
\usepackage{bbm}
\usepackage{mathtools}

\def\be{\begin{equation}}
\def\ee{\end{equation}}

\def\Tr{{\rm Tr}}

\def\IC{\mathbb{C}}

\def\IM{\mathbb{M}}

\def\IP{\mathbb{P}}
\def\IQ{\mathbb{Q}}
\def\IR{{\mathbb{R}}}

\def\IZ{{\mathbb{Z}}}

\def\CB{{\cal B}}

\def\CM {{\cal M}}

\def\CN {{\cal N}}
\def\CR {{\cal R}}

\def\CF {{\cal F}}

\def\CL {{\cal L}}
\def\CV {{\cal V}}
\def\CW {{\cal W}}

\def\CE {{\cal E}}

\def\CH {{\cal H}}
\def\CI {{{\cal I}}}
\def\CB {{\cal B}}
\def\CQ {{\cal Q}}
\def\CS {{\cal S}}
\def\CT {{\cal T}}
\def\CU {{\cal U}}

\def\half{\frac{1}{2}}

\renewcommand{\Im}{{\rm Im }}
\renewcommand{\Re}{{\rm Re }}

\def\one{{\hbox{ 1\kern-.8mm l}}}

\def\vol{{\rm vol\,}}
\def\p{\partial}

\def\be{\bar{e}}

\def\half{\frac{1}{2}}

\def\Re{{\rm Re}}

\def\half{\frac{1}{2}}

\def\Re{{\rm Re}}

\def\be{ \begin{equation} }
\def\ee{ \end{equation}}

\def\fa{\mathfrak{a}}

\def\fc{\mathfrak{c}}

\def\fr{\mathfrak{r}}
\def\fs{\mathfrak{s}}
\def\ft{\mathfrak{t}}

\def\fr{\mathfrak{r}}
\def\fs{\mathfrak{s}}
\def\ft{\mathfrak{t}}
\def\fu{\mathfrak{u}}

\def\fw{\mathfrak{w}}

\def\fB{\mathfrak{B}}

\def\fE{\mathfrak{E}}
\def\fI{\mathfrak{I}}

\def\fS{\mathfrak{S}}
\def\fT{\mathfrak{T}}

\def\fV{\mathfrak{V}}

\def\Tr{{\rm Tr}}

\def\IC{\mathbb{C}}

\def\IM{\mathbb{M}}

\def\IP{\mathbb{P}}
\def\IQ{\mathbb{Q}}
\def\IR{{\mathbb{R}}}

\def\IV{{\mathbb{V}}}

\def\IZ{{\mathbb{Z}}}

\def\I{{\rm i}}
\def\afty{{$A_{\infty}$}}

\newcommand\fro{{\overline{\underline{\Omega}}}}

\def\Rvtx{ R^{\rm int}}

\def\fVac{ { \fV \fa\fc} }
\def\IntfcTimes{ \boxtimes }
\def\Hom{{\rm Hom}}
\def\Hop{{\rm Hop}}
\def\fId{\mathfrak{I}\mathfrak{d}}
\def\Id{\boldsymbol{\mathrm{Id}}}
\def\Map{{\rm Map}}
\def\SpecGen{ \widehat{R} }

\def\d{{\mathrm d}}
\def\D{{\mathcal D}}

\title{An Introduction To The Web-Based Formalism}

\author{Davide Gaiotto,$^1$  Gregory W. Moore,$^2$  and Edward Witten$^3$\\
$^1$Perimeter Institute for Theoretical Physics\\
31 Caroline Street North, ON N2L 2Y5, Canada\\
 $^2$ NHETC and Department of Physics and Astronomy,
Rutgers University,\\
Piscataway, NJ 08855--0849, USA\\
$^3$ School of Natural Sciences, Institute for Advanced Study, \\
Princeton, NJ 08540, USA\\
\\
{\rm dgaiotto@gmail.com, gmoore@physics.rutgers.edu, witten@ias.edu} }

\abstract{This paper summarizes our rather lengthy paper,
``Algebra of the Infrared: String Field Theoretic Structures
 in Massive $\CN=(2,2)$  Field Theory In Two Dimensions,''
and is meant to be an informal, yet detailed, introduction and
summary of that larger work. }

\begin{document}

\section{Background And Motivation}\label{sec:Background}

\subsection{Introduction}

This paper summarizes our rather lengthy paper,
``Algebra of the Infrared: String Field Theoretic Structures
 in Massive $\CN=(2,2)$  Field Theory In Two Dimensions,''
 henceforth cited as \cite{Gaiotto:2015aoa}. The present
 paper  is meant to be a very informal, yet detailed, introduction and
summary of that larger work. See \cite{Gaiotto:2015aoa} for more references and
more background. The reader who finds our presentation to be too
telegraphic at some points  is encouraged to consult the
main text  for a more leisurely account.

\subsection{Goals And Motivation}\label{subsec:Goals}

Let $X$ be a K\"ahler manifold, and $W: X\rightarrow \IC$ a holomorphic
Morse function. To this data physicists associate a ``Landau-Ginzburg (LG) model.'' It is
closely related to the Fukaya-Seidel (FS) category. The goals of this introduction
are:

\begin{enumerate}

\item   To construct an \afty- category of branes in this model,
using only data ``visible at long distances'' - that is, only data about
BPS solitons and their interactions. This is the ``web-based formalism.''

\item  To explain how the ``web-based'' construction of an \afty-category
of branes is related to the  FS category.

\item To construct an \afty\  2-category of theories, interfaces, and boundary operators.

\item  To show how these interfaces categorify the wall-crossing formula
for BPS solitons  as well as the wall-crossing formulae for framed BPS states.

\end{enumerate}

Two of the motivations for the detailed construction of interfaces are
the nonabelianization map of Hitchin systems that arises in theories of
class S \cite{Gaiotto:2012rg}, and the application of supersymmetric gauge theory to knot
homology \cite{Witten:2011zz}.  We will return to these
motivations briefly in Sections \S \ref{subsubsec:Cat-S-Wall} and
\S \ref{subsec:KnotHomology}, respectively. These applications
have only been partially worked out and
remain interesting topics for further research. They are
described in more detail in Sections \S 18.2 and \S  18.4 of \cite{Gaiotto:2015aoa},
respectively.

\subsection{A Review Of Landau-Ginzburg Models}

To warm up, let us review some well-known facts about  Landau-Ginzburg theory
in two dimensions. We want to understand
the groundstates of the model in various geometries with
various boundary conditions.  We approach
the subject from the viewpoint of Morse theory.

\subsubsection{Supersymmetric Quantum Mechanics And Morse Theory}

From a physicist's point of view Morse theory is the theory
of the computation of groundstates in supersymmetric quantum
mechanics (SQM) \cite{Witten:1982im}.
Recall that in SQM we have a particle moving on
a Riemannian manifold $q: \IR \to M$ together with
a real Morse function $h:M \to \IR$ and we consider
the (Euclidean) action

\be
S_{SQM} = \int dt \left( \half \vert \dot q \vert^2 + \half \vert dh \vert^2 + \cdots \right)
\ee
There is a uniquely determined perturbative vacuum
 $\Psi(p_i)$ associated  to each critical point  $p_i$ of $h$.
 True vacua are linear combinations of the $\Psi(p_i)$. How do we find them?

To find the true vacua we introduce the   MSW (``Morse-Smale-Witten'') complex
generated by the perturbative ground states
\be
\IM = \oplus_{ p_i: dh(p_i)=0} \IZ \cdot \Psi(p_i).
\ee
The complex is graded by the Fermion number operator $\CF$, whose value on $\Psi(p_i)$ is:
\be
f = \half (n_-  - n_+ )
\ee
where $n_\pm$ is the number of $\pm $ eigenvalues of the Hessian.
The   matrix elements of the differential $Q$ are obtained by
counting the number of solutions to the instanton equation:
\be
\frac{dq}{d\tau} = \nabla h
\ee
which have no reduced moduli and interpolate between two critical points.
By ``counting'' we always mean ``counting with signs determined by certain orientations.''
The space of true ground states is the cohomology $H^*(\IM, Q)$ of the MSW complex.

\subsubsection{Landau-Ginzburg Models From Supersymmetric Quantum Mechanics}

Now, to formulate LG models,  we apply the SQM formulation of Morse theory
to the case where the target manifold $M$ of the SQM is a space of maps $D\to X$,
and $D$ is a one-dimensional manifold, possibly with boundary:
\be
M = \Map(D \to X ).
\ee
The real SQM Morse function is
\be\label{eq:MorseFun}
h = -  \int_D   \left( \phi^*(\lambda) - \frac{1}{2} \Re(\zeta^{-1} W)dx \right).
\ee
Here $\zeta$ is a phase.
For simplicity we assume that the K\"ahler manifold is exact and
choose a trivialization of the symplectic form $ \omega = d \lambda$.
Recall that $W: X \to \IC$ is a holomorphic Morse function. This means
that at the critical points where $dW(\phi_i)=0$ the Hessian $W''$ is
nondegenerate.   If we work out the
SQM action   we get a $1+1$ dimensional
field theory. The bosonic terms in the action are
\be
\int_{D\times \IR}  \half \vert d\phi\vert^2 + \half \vert \nabla W \vert^2 + \cdots
\ee

The perturbative groundstates, from the SQM viewpoint, are solutions of
 $\delta h=0$. This equation is equivalent to   the \emph{$\zeta$-soliton equation}:
\be\label{eq:LG-flow}
 \frac{\d}{\d x} \phi^I = g^{I\bar J}\frac{\I \zeta}{2}
\frac{\p \bar W}{\p \bar \phi^{\bar J}}.
\ee
(Later we will find it useful to note that the $\zeta$-soliton equation is
equivalent to both upwards gradient flow with potential $\Im(\zeta^{-1} W)$
as well as   Hamiltonian flow with Hamiltonian $\Re(\zeta^{-1} W)$.)

One solution of \eqref{eq:LG-flow} is given by constant field configuration
\be
\phi(x,t) = \phi_i \in \IV
\ee
where $\IV$ denotes the set of critical points of $W$.  If these are compatible with
the boundary conditions they turn out to be
true vacua, and they are massive vacua if $W$ is Morse. However, it is possible to
consider boundary conditions so that  \eqref{eq:LG-flow} does not admit solutions where  $\phi$
is a constant.
These are called \emph{soliton} solutions, and we turn to them next.

\subsubsection{Solitons On The Real Line}

Now suppose $D = \IR$. We choose boundary conditions of finite energy:

\be\label{eq:left-infty-bc}
\lim_{x\to -\infty} \phi = \phi_i
\ee
\be\label{eq:right-infty-bc}
 \lim_{x\to + \infty } \phi= \phi_j
\ee
where $\phi_i, \phi_j \in \IV$ and  $\phi_i \not= \phi_j$.
What is the MSW complex in this case?

It is a standard fact that solutions to \eqref{eq:LG-flow} project to straight lines of
slope $\I\zeta$ in the complex $W$-plane. Therefore, there is
no solution for generic $\zeta$. There can only be a solution for
\be\label{hopeful}
\I \zeta =
\I \zeta_{ji} := \frac{W_j - W_i}{\vert W_j - W_i \vert}
\ee
in which case a solution projects in the $W$-plane to a line
segment between the critical values $W_i$ and $W_j$ of $W$.

Now, to describe the solutions we introduce the notion of a
\emph{Lefshetz thimble}. This is the maximal (i.e. maximal dimension)
Lagrangian subspace of $X$ defined by the  inverse image in $X$ of all solutions
to \eqref{eq:LG-flow} satisfying the boundary condition \eqref{eq:left-infty-bc},
(respectively \eqref{eq:right-infty-bc}). Solutions satisfying \eqref{eq:left-infty-bc}
are known as left-Lefshetz thimbles and those satisfying \eqref{eq:right-infty-bc}
are known as right-Lefshetz thimbles.  Evidently, a soliton solution for $D=\IR$ must simultaneously
be in a left and a right Lefshetz thimble, and hence sits in the intersection of
the two.  We assume that the left- and right- Lefshetz thimbles intersect
transversally in the fiber over a regular value of $W$ on the
line segment $[W_i,W_j]$. Denote this set of intersections by $\CS_{ij}$.
There will be a finite
number of classical solitons, one for each intersection point
$p\in \CS_{ij}$. The MSW complex turns out to be:
\be\label{eq:MorseComplexR}
\IM_{ij}= \oplus_{p\in \CS_{ij} } \left( \IZ \Psi^f_{ij}(p)\oplus \IZ \Psi^{f+1}_{ij}(p) \right).
\ee
The grading of the complex is given by the fermion number of the perturbative ground state.
This turns out to be given by $f$ or $f+1$ for the two generators above where
\begin{equation}\label{zelbor} f=  -\frac{\eta(\D+\varepsilon)}{2}.\end{equation}
Here $\D$ is the Dirac operator obtained by linearizing the $\zeta$-soliton equation
\eqref{eq:LG-flow}, $\varepsilon$ is small and positive, and $\eta(\D)$ denotes the
\emph{eta invariant} of Atiyah, Patodi, and Singer.
\footnote{Roughly speaking, the $\eta$ invariant of a self-adjoint operator
is a regularized version of the sum of signs of the eigenvalues of the operator. In general
it is not integer, but the difference of eta invariants for different $ij$ solitons will be
an integer. This is why the MSW complex is graded by a $\IZ$-torsor rather than by $\IZ$.
There is a tricky point here: In the algebraic manipulations below
it is important to use the Koszul rule, a rule that only makes sense when there
is an integral grading. One needs to write $f= f_i - f_j + n_{ij}$, where $n_{ij}$ is
integral, and remove the $f_i$ by a kind of ``gauge transformation'' of the wavefunctions.
Then the complex is $\IZ$-graded.}
We can now introduce the \emph{BPS soliton degeneracies} \cite{Cecotti:1992qh}:
\be
\mu_{ij}:= - \Tr_{\IM_{ij}}  {\rm \textbf{F}} e^{\I \pi {\rm \textbf{F}}}
\ee
where ${\rm \textbf{F}}$ is the Fermion number operator, taking values $f$ and $f+1$ on the
perturbative groundstates  $\Psi^f_{ij}(p)$ and $\Psi^{f+1}_{ij}(p)$, respectively.
The degeneracies $\mu_{ij}$  will show up in  \S \ref{subsec:Cat-Muij}
and again in  \S \ref{sec:Interface-WC} when we discuss wall-crossing. We can already note that,
in some sense, $\IM_{ij}$ has ``categorified the 2d BPS degeneracies.''

\begin{figure}[h]
\centering
\includegraphics[scale=0.50,angle=0,trim=0 0 0 0]{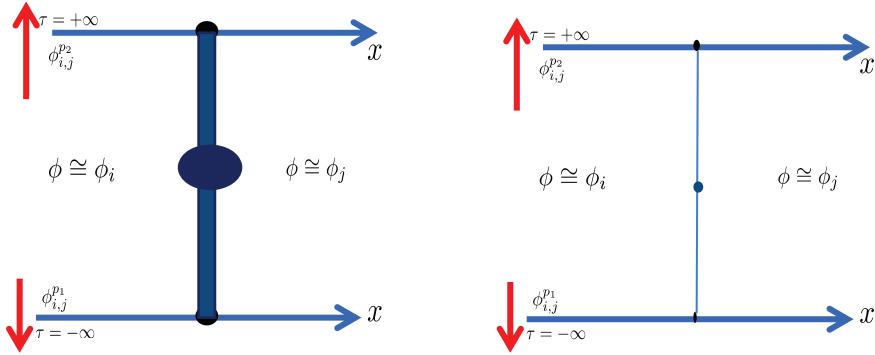}
\caption{\small Left: An instanton configuration contributing to the
differential on the   MSW complex. The black regions
indicate the locus where the field $\phi(x,\tau)$ varies vary significantly from the
vacuum configurations $\phi_i$ or $\phi_j$. The length scale here is $\ell_W$,
set by the superpotential $W$.   Right: Viewed from a large
distance compared to the length scale $\ell_W$ the instanton looks like a
straight line $x=x_0$, where the vacuum changes discontinuously from vacuum $\phi_i$
to $\phi_j$. The nontrivial $\tau$-dependence of the instanton configuration,
interpolating from a   soliton $p_1$ to another soliton $p_2$ has been contracted to
a single vertex located at $\tau = \tau_0$. This vertical line with a single vertex on it
is the first example of a ``web'' in the web formalism. }
\label{fig:INSTANTON-ON-R}
\end{figure}

The differential on $\IM_{ij}$ is given by following the SQM paradigm: We count instantons.
In the present case the SQM instantons are   solutions to the \emph{$\zeta$-instanton equation}:
\be\label{eq:LG-INST}
\left(\frac{\p  }{\p x} +\I  \frac{\p }{\p \tau} \right)\phi^I = \frac{\I \zeta}{2}g^{I\bar J} \frac{\p \bar W}{\p \bar\phi{}^{\bar J}},
\ee
with boundary conditions illustrated in Figure \ref{fig:INSTANTON-ON-R}.

Written out the boundary conditions for the $\zeta$-instanton equation are:
\be \label{bcond}
\lim_{x\to -\infty}\phi(x,\tau)=\phi_i \qquad \lim_{x\to +\infty}\phi(x,\tau)=\phi_j
\ee
\be\label{ccond}
\lim_{\tau\to - \infty} \phi(x,\tau) = \phi^{p_1}_{ij}(x) \qquad
\lim_{\tau\to + \infty} \phi(x,\tau) = \phi^{p_2}_{ij}(x).
\ee
Following the rules of SQM, the matrix elements of the differential are obtained
by counting the solutions with no reduced moduli, (i.e. the solutions
with  two moduli).

\textbf{Remarks:}

\begin{enumerate}

\item   The complex \eqref{eq:MorseComplexR}
is not a standard mathematical Morse theory complex:
$h$ is degenerate because of translation invariance. The critical
set is $\IR$, parametrizing the ``center of mass'' of the soliton.
 But we sum neither the cohomology  nor the compactly
supported cohomology of this critical set, as one would do in standard Morse theory. Rather, we
attach a certain Clifford module to each critical locus.
Physically this arises from the   quantization
of the ``collective coordinates'' associated with the center of mass of the
soliton. The module has rank two. That is why each classical soliton
contributes two perturbative groundstates in equation \eqref{eq:MorseComplexR}.
In equation \eqref{eq:Rij-WebRep} below we have factored out this center of mass degree of
freedom and hence each soliton leads to just one perturbative groundstate.

\item Supersymmetric quantum mechanics has two supersymmetries
satsifying $\{ Q, \bar Q \} = 2H$. When the spatial domain is $D=\IR$
there are more symmetries in the problem, such as translational symmetry
along $\IR$,  not manifest from the general SQM
viewpoint. Consequently, when the spatial domain is $\IR$ the LG model has (2,2) supersymmetry:
\be\label{eq:22susy}
\begin{split}
 \{Q_+,\bar Q_+\} = H+P \qquad &  \qquad  \{Q_+,  Q_-\} =\bar Z \\
  \{Q_-,\bar Q_-\}  =H-P \qquad &  \qquad   \{\bar Q_+,\bar Q_-\} =  Z. \\
\end{split}
\ee
The supersymmetries of the SQM are of the form
\be \label{manifest} \CQ_{\zeta}:=Q_- -\zeta^{-1}\bar Q_+,~~\bar{\CQ}_{\zeta} :=\bar Q_- -\zeta   Q_+. \ee
The vacua of the model $\phi_i\in \IV $ preserve four supersymmetries.
The $\zeta$-soliton equation is the  $\CQ_{\zeta}$ (or $\bar \CQ_{\zeta}$)-fixed point equation for 
stationary classical field configurations. Solutions of these equations preserve two out of the four
supersymmetries. The $\zeta$-instanton equation is an equation for the theory in Euclidean signature 
and preserves only one supersymmetry, namely $\CQ_{\zeta}$.  When $D$ is a half-line or an interval, with suitable boundary
conditions, only the two-dimensional supersymmetry algebra generated by $ \CQ_{\zeta}$ and $\bar{\CQ}_{\zeta}$ will be 
preserved. 

\item Now comes an important physics point: The theory is \emph{massive} with a length scale
$\ell_W$ corresponding to the inverse of the lightest soliton mass. Physical correlations
should decay exponentially beyond that scale. We can picture the solitons and
instantons as in Figure \ref{fig:INSTANTON-ON-R}.

\item The $\zeta$-instanton equation has also appeared in the literature on the relation of matrix models
and Landau-Ginzburg models \cite{WittenAlgebraicGeometry,Fan:2007ba}. It also appears
 in the literature on BPS domain walls in four-dimensional supersymmetric
theories \cite{Carroll:1999wr,Gibbons:1999np}. There are even some exact solutions available in the literature
\cite{Oda:1999az}.

\end{enumerate}

\subsection{LG Models On A Half-Plane And The Strip}

\subsubsection{Boundary Conditions}

If $D$ has a left-boundary $x_\ell \leq x$ or a right boundary $x\leq x_r$
at finite distance then we need to put boundary conditions to get a good
Morse theory, or QFT.

\begin{enumerate}

\item At $x=x_\ell,  x_r$, the boundary value
$\phi^\p$ must be valued in a maximal Lagrangian submanifold $\CL_\ell, \CL_r$ of $X$
in order to have elliptic boundary conditions for the Dirac equation on the fermions.

\item The theory is simplest when the Lagrangian submanifolds are exact: $\iota^*(\lambda) = dk$, for a single-valued
function $k$. Indeed, the Morse function \eqref{eq:MorseFun} is replaced
by $h \to h \pm k(\phi^\p)$, where the sign is for the negative/positive half-plane, respectively.
Note that $k$ can thus be interpreted physically as a boundary superpotential.

\end{enumerate}

We are certainly interested in $X$ which is noncompact (since we want $W$ to be nontrivial)
and we are typically interested in noncompact Lagrangians. Now, we want  to have well-defined
spaces of quantum states on an interval $\CH_{\CL_{\ell}, \CL_{r}}$, invariant under separate
Hamiltonian symplectomorphisms of the left and right branes. (These are mirror dual to gauge
transformations on the branes of the B-model.)

The generators of the MSW complex in this case can be identified with the intersection points
\be
\CL_\ell^{(\Delta x)} \cap \CL_r
\ee
where we regard the $\zeta$-soliton equation \eqref{eq:LG-flow} as a flow in $x$
and $\CL^{(\Delta x)}$ means the flow has been applied for a range $(\Delta x)$.

\begin{figure}[h]
\centering
   \includegraphics[width=2.2in]{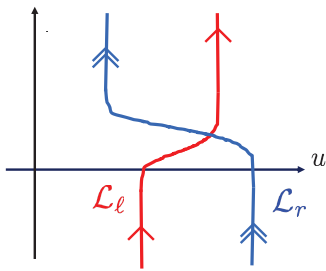}
\caption{\small A pair of Lagrangian submanifolds $\CL_\ell$, $\L_r$ embedded in the $u-v$ plane.  $\CL_\ell$ and $\CL_r$ intersect at the one
point indicated.  $u$ is plotted horizontally and we assume that $\CL_\ell$, $\CL_r$
are embedded in the half-plane $u>0$.}
 \label{fig:Lagrangians}
\end{figure}

But now there is a problem: Intersection points can go to infinity as the length of the
interval is changed (or if independent Hamiltonian symplectomorphisms are applied to left
and right branes). As an example, consider  $\zeta^{-1} W = \I \phi^2 $ and consider the candidate left and
right branes shown in Figure \ref{fig:Lagrangians}. We regard the $\zeta$-soliton equation as
a flow in $x$, and if $\phi = u + \I v$ is the decomposition into real and imaginary parts
then
\be
\p_x u = u  \qquad \p_x v = - v
\ee
Therefore, the flow in $x$ of $\CL_\ell$ will not intersect $\CL_r$ for sufficiently large $x$.
Therefore there will be supersymmetric states for small width of $[x_\ell, x_r]$ but none for large
width of $[x_\ell, x_r]$. This is potentially an interesting feature for a physicist studying
supersymmetry breaking, but it is a bug for the kind of ``partial topological field theory''
we are studying.

\begin{figure}[h]
\centering
\includegraphics[width=1.5in]{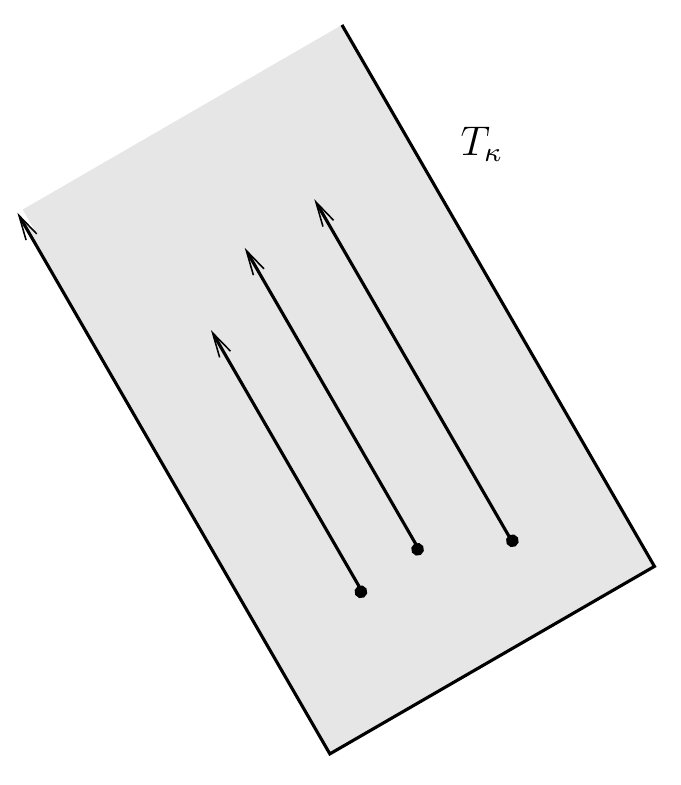}
\caption{\small  The rays in the complex $W$-plane that start at critical points and all run in the $\i\kappa$ direction fit into the semi-infinite strip $T_\kappa$, which is shown as a shaded region. }\label{manyrays}
\end{figure}

In \cite{Gaiotto:2015aoa} we find that there are \emph{two} distinct criteria we could impose
on the allowed Lagrangians to avoid the above problem. One solution is to restrict the left
and right branes to be positively and negatively $W$-dominated, respectively. A
brane supported on $\CL$ is positively (negatively) $W$-dominated if $\Im(\zeta^{-1} W) \to \pm \infty$
as $\phi$ goes to infinity along $\CL$. Alternatively, one can restrict
 the Lagrangians to be   \emph{Branes of class $T_{\kappa}$}:
Choose a phase $\kappa\not= \pm \zeta$,
and constants $c,c'$. The precise choices don't matter too much, although which
component of the circle $\kappa$ sits in is significant. Branes of class $T_{\kappa}$
are based on Lagrangians which project under $W$ to a semi-infinite rectangle in
the $W$-plane:
\be\label{zelbo}
\begin{split} |\mathrm{Re}\,(\kappa^{-1}W)| & \leq  c\cr
 \mathrm{Im}\,(\kappa^{-1}W)& \geq c',\\ \end{split}
\ee
as in Figure \ref{manyrays}. In the second approach branes on both the left
and right boundaries are taken to be in class $T_{\kappa}$. Now,   under the $x$-flow of the
$\zeta$-soliton equation we have
\be\label{eq:x-flow}
\frac{d}{dx} \mathrm{Re}\,(\kappa^{-1}W) = -\half \{ \Re(\zeta^{-1} W), \Re(\kappa^{-1} W) \}= \frac{1}{4} \Im(\frac{\zeta}{\kappa})
 \vert dW \vert^2
\ee
Then, points at infinity flow very fast out of the rectangle and hence intersection points $\CL_\ell^{(\Delta x)} \cap \CL_r$
always sit in a bounded region and cannot escape to infinity.

\subsubsection{LG Ground States On A Half-Line}

Now we consider the theory on the positive half-plane.
We choose $\zeta$ so that it does \emph{not} coincide with any of the $\zeta_{ij}$ defining
the solitons for $D=\IR$. What are the groundstates preserving $\CQ_\zeta$ supersymmetry?

The MSW complex $\IM_{\CL_{\ell}, j}$  is generated by the $\zeta$-solitons on the half-plane satisfying the
above boundary conditions.
The grading on the complex is again given by fermion number but finding a formula for the
fermion number is a little nontrivial. We only know how to describe it when $X$ is Calabi-Yau.
In this case case we define
\begin{equation}\label{dommy}e^{\i\vartheta}=\frac{\mathbf{vol}}{\Omega|_{\CL}} \end{equation}
(where $\Omega$ trivializes $K_X$ and is normalized so that $\Omega \bar \Omega$ is the volume
form on $X$) and we need to be able to define a single-valued logarithm $\vartheta$. (That is,
the Maslov index must vanish.) In this case
we define the fermion number (on the interval) to be:
\begin{equation}\label{cgt} f= -\frac{1}{2}\eta(\D ) -2\frac{\varphi_r-\varphi_\ell}{2\pi}.  \end{equation}
where $\varphi  = \vartheta(\phi^\p)$.
On a half-line we drop $\varphi_r$ or $\varphi_\ell$ as appropriate.

\begin{figure}[h]
\centering
\includegraphics[scale=0.37,angle=0,trim=0 0 0 0]{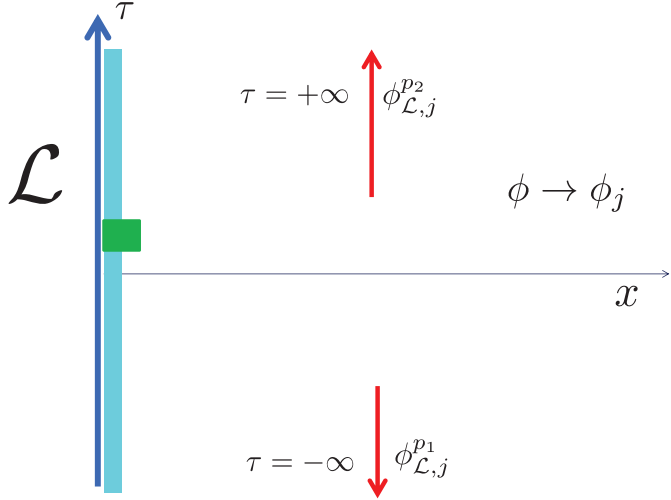}
\caption{\small An instanton in the complex $\IM_{\CL,j}$.  The solitons corresponding to
$p_1,p_2 \in \CL\cap R^\zeta_j$, where $R^\zeta_j$ is the right Lefshetz thimble,
 are exponentially close to the vacuum $\phi_j$ except
for a small region, shown in turquoise,  of width $\ell_W$. In addition, the instanton
transitions from one soliton to another in a time interval of length $\ell_W$,
indicated by the green square. At large distances the green square becomes the
$0$-valent vertex used in   half-plane webs. }
\label{fig:HALFPLANE-INSTANTON-1}
\end{figure}

The differential on the complex is given by counting $\zeta$-instantons.
The picture of the instantons on the half-plane is shown in Figure \ref{fig:HALFPLANE-INSTANTON-1}

\subsubsection{LG  Ground States On The Strip}

The story on the strip is very similar to that on the half-plane, but
there is an interesting wrinkle that provides a nice example
where naive categorification of formulae for BPS degeneracies fails:
We consider the LG theory on $\IR \times [x_\ell, x_r]$.
When $\vert x_r - x_{\ell} \vert \gg \ell_W$ the $\zeta$-solitons must
nearly ``factorize'' so there is a natural isomorphism:
\be\label{eq:appxt-complex-ii}
 \IM_{\CL_\ell, \CL_r} \cong \oplus_{i\in
 \IV}  \IM_{\CL_\ell,i} \otimes \IM_{i, \CL_r}.
\ee
So if we define the BPS degeneracy of the half-line solitons:
\be
\mu_{\CL,i} := {\Tr}_{\IM_{\CL,i}} e^{i \pi {\rm \textbf{F}} },
\ee
and similarly define $\mu_{i, \CL}$ and $\mu_{\CL_\ell, \CL_r}$,
then the Euler-Poincar\'e principle guarantees
\be\label{eq:WittenIndexFactorize}
\mu_{\CL_\ell, \CL_r} = \sum_{i \in \IV} \mu_{\CL_\ell,i} \mu_{i, \CL_r}.
\ee
Now, the naive categorification would state:
\be\label{wrong}
H^*( \IM_{\CL_\ell, \CL_r})\overset{?}{\cong} \oplus_{i\in \IV}  H^*(\IM_{\CL_\ell,i}) \otimes H^*(\IM_{i, \CL_r}).
\ee
Here we have used the natural differential on the tensor-product complex. It corresponds to the
$\zeta$-instantons of Figure \ref{fig:NaiveStripDifferential}:
\begin{figure}[h]
\centering
\includegraphics[scale=0.35,angle=0,trim=0 0 0 0]{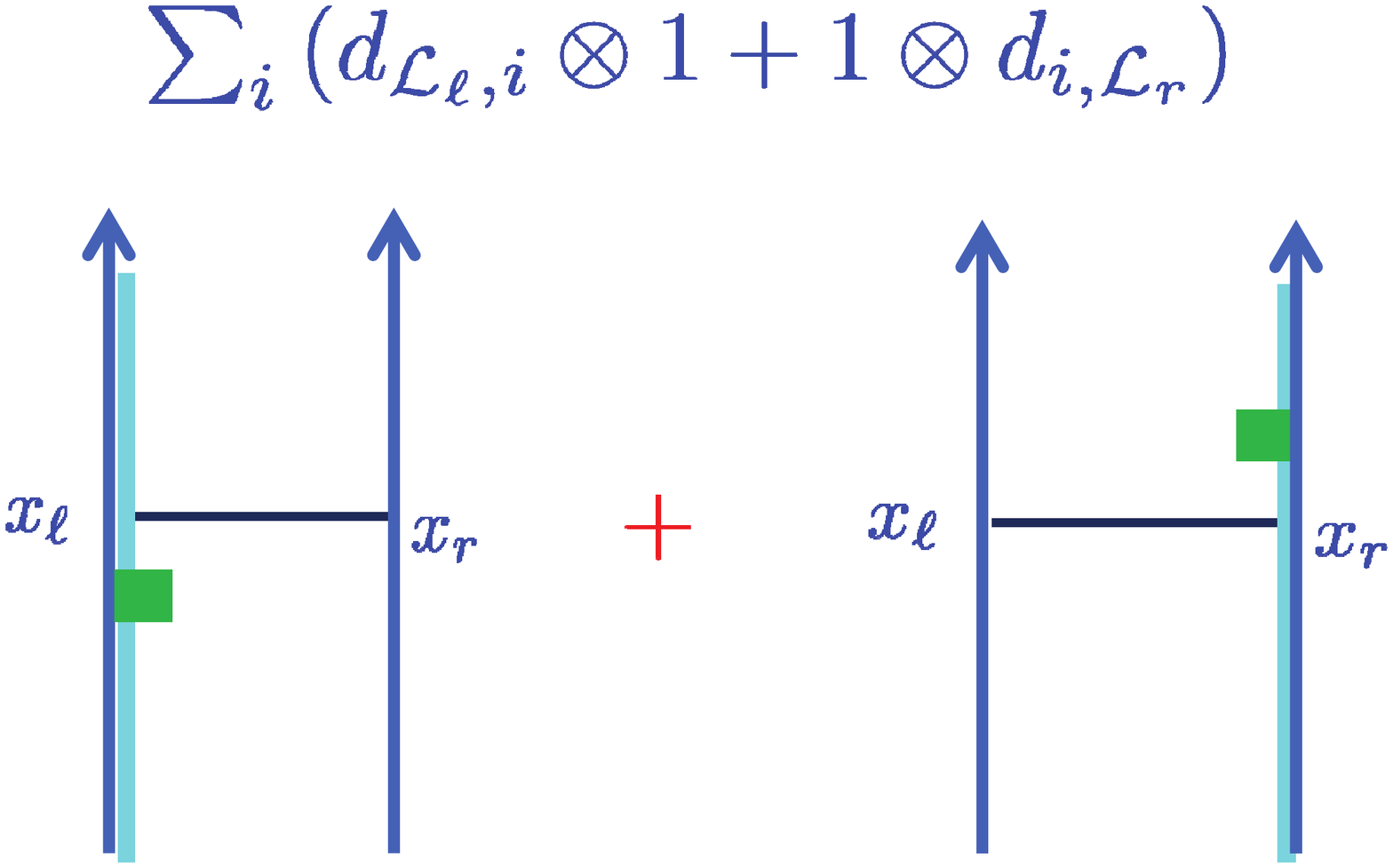}
\caption{\small Naive differential on the strip.
   }
\label{fig:NaiveStripDifferential}
\end{figure}

\begin{figure}[h]
\centering
\includegraphics[scale=0.35,angle=0,trim=0 0 0 0]{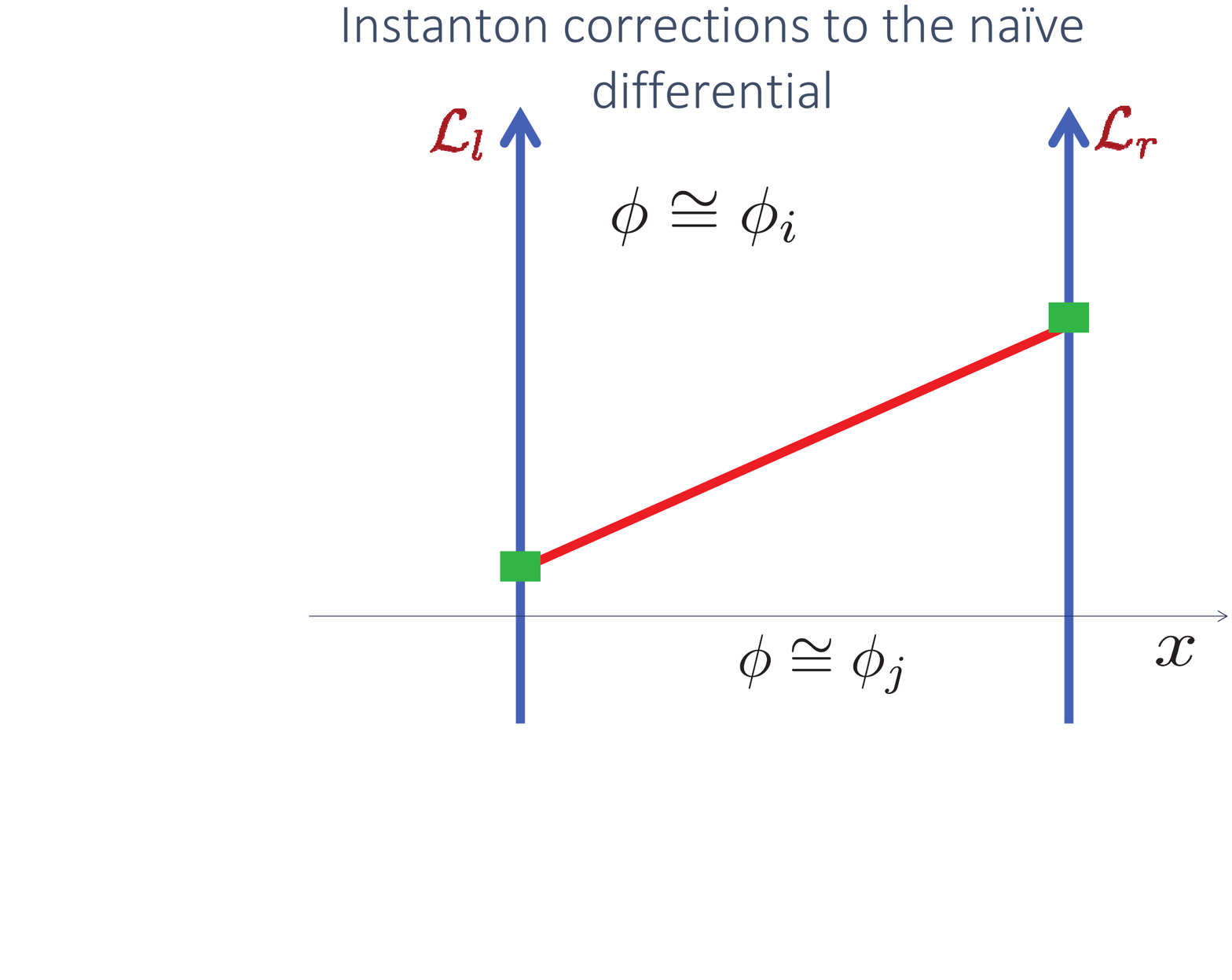}
\caption{\small An instanton correction to the naive differential on the strip.
   }
\label{fig:StripInstanton}
\end{figure}

As we will see, equation \eqref{wrong} is wrong. The reason is that there are other $\zeta$-instantons
which also contribute to the physically correct differential.
 One example is a $\zeta$-instanton that looks like Figure \ref{fig:StripInstanton}.
We will interpret this figure more precisely at the end of \S \ref{subsubsec:BoostSol}.

\subsection{A Physicist's View Of The Fukaya-Seidel Category  }

Finally, we sketch the Fukaya-Seidel (FS) category, at least the way a
physicist would approach it (after benefiting from exposure to mathematical thinking on this topic).\footnote{We thank Nick Sheridan for many useful discussions about
the mathematical approaches to the FS category.}

\begin{figure}[h]
 \centering
 \includegraphics[scale=0.55,angle=0,trim=0 0 0 0]{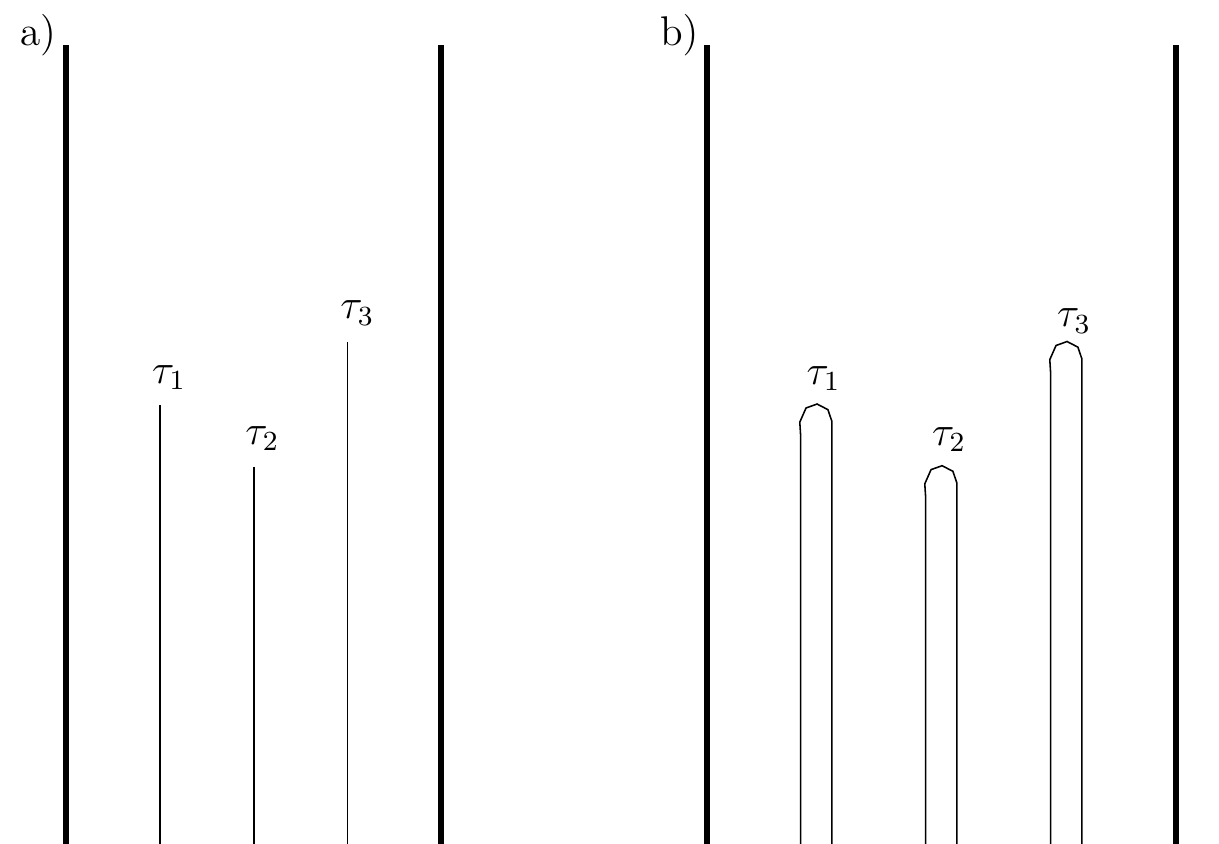}
\caption{\small A picture of the worldsheet of
  $n$ open strings all of width $w$ coming in from the past ($\tau=-\infty$) with a single open string of width $nw$
going out to the future ($\tau=+\infty$), familiar from the light-cone gauge formulation of
string interactions. (a)  There are $n-1$ values of $\tau$ at which two open strings combine to one.  The
 linearly independent differences between these critical
values of $\tau$ are the $n-2$ real moduli of this worldsheet.  (b)  The picture in (a) can be slightly modified in this fashion  so that the
worldsheet is smooth.   The moduli are still the differences between the critical values of $\tau$.} \label{fig:Worldsheets}
\end{figure}

Fix $\zeta$. Our objects will be branes based on Lagrangians in class $T_{\kappa}$,
where $\kappa$ is in one of the two components of $U(1) - \{ \pm \zeta \}$. Up to
$A_\infty$ equivalence the category should only depend on the choice of component.
The morphism space is the MSW complex $\IM_{\CL_\ell, \CL_r}$ generated by
solutions of the $\zeta$-soliton equation.
Then, to compute the differential $M_1$, we count $\zeta$-instantons with one-dimensional
moduli space. (That is, zero-dimensional reduced moduli space.)
To compute the higher \afty-products we follow the example of open string field theory in
light-cone gauge. We divide up the interval into equal length subintervals
and consider the diagram in Figure \ref{fig:Worldsheets}.
Finally, we have to integrate over the moduli -  the relative positions of the joining times.
When the fermion numbers of the incoming and outgoing states are such
that the amplitude is not trivially zero the expected dimension of the moduli
space will be zero, and in fact the solutions will only exist for a finite
set of critical values $\tau_i - \tau_{i+1}$ where the strings join.
The   amplitude is obtained by counting
over the finite set of solutions to the $\zeta$-instanton equation.

  The \afty-category we have sketched above
is  not precisely what we one finds in the literature.
(See, for example \cite{SeidelBook}.) Our understanding from experts  is that something roughly along the lines of what we have written using the $\zeta$-instanton
equations is understood to be the natural conceptual framework for defining the FS category, and that proofs along these
lines will materialize in the literature in due course.

\section{The Web Formalism}\label{sec:WebFormalism}

\subsection{Boosted Solitons And $\zeta$-Webs}

Now we would like to interpret more precisely the meaning of Figure \ref{fig:StripInstanton}.

\subsubsection{Boosted Solitons}\label{subsubsec:BoostSol}

Recall that $\zeta$-instantons satisfy
\be\label{eq:LG-INST-p}
\left(\frac{\p  }{\p x} +\I  \frac{\p }{\p \tau} \right)\phi^I = \frac{\I \zeta}{2}g^{I\bar J} \frac{\p \bar W}{\p \bar\phi{}^{\bar J}},
\ee
and we are interested in solutions for arbitrary phase $\zeta$.
Recall too that $\zeta$-solitons on $D=\IR$ satisfy
\be\label{eq:LG-flow-p}
 \frac{\d}{\d x} \phi^I = g^{I\bar J}\frac{\I \zeta}{2}
\frac{\p \bar W}{\p \bar \phi^{\bar J}}.
\ee
Moreover,  with boundary conditions $(\phi_i,\phi_j)$ at $x=-\infty,+\infty$ solutions only exist for
special phases $\I \zeta_{ji}$ given by the phase of the difference of critical values
$W_j - W_i$.

\begin{figure}[h]
\centering
\includegraphics[scale=0.4,angle=0,trim=0 0 0 0]{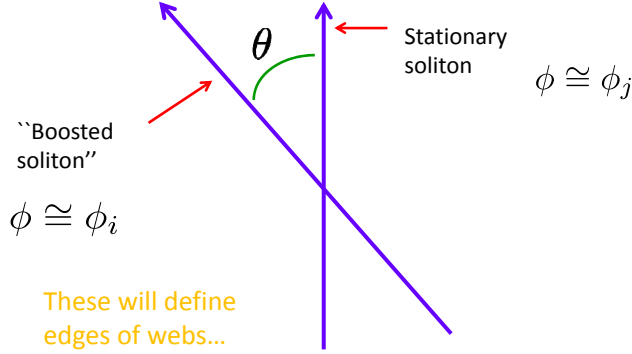}
\caption{\small The boosted soliton. A short computation show that the ``worldline'' is
parallel to the complex number $z_{ij}:= z_i - z_j$ where $z_i = \zeta \bar W_i$.  }
\label{fig:BoostedSolitonFlorida}
\end{figure}

We can nevertheless use solitons of type $ij$ to produce solutions of the $\zeta$-instanton equation
on the Euclidean plane by taking the ansatz:
\be
\phi_{ij}^{\rm boosted}(x,\tau) := \phi_{ij}^{\rm soliton}(\cos\theta x + \sin\theta \tau).
\ee
Since
\be\label{eq:boosted-soliton}
\left(\frac{\p  }{\p x} + \I \frac{\p }{\p \tau} \right)\phi_{ij}^{{\rm boosted},I}(x,\tau)
=  \frac{\I e^{\I \theta}\zeta_{ji} }{2}g^{I\bar J}\partial_{\bar J}\bar W(\phi_{ij}^{{\rm boosted}})
\ee
it follows that if we choose the rotation $\theta$ so that
\be\label{eq:xi-to-zeta}
e^{\I \theta}  \zeta_{ji} = \zeta
\ee
then we obtain a solution to the $\zeta$-instanton equation.
We call such solutions to the $\zeta$-instanton equation \emph{boosted solitons}.
A short computation shows that the ``worldline'' (i.e. the region where the solution is not
exponential close to one of the vacua $\phi_i$ or $\phi_j$) is
parallel to the complex number $z_{ij}:= z_i - z_j$ where $z_i = \zeta \bar W_i$.
See Figure \ref{fig:BoostedSolitonFlorida}.

Now we can start to interpret the ``extra''   $\zeta$-instanton
illustrated in Figure \ref{fig:StripInstanton}. The idea is that
if the width of the interval is much larger than $\ell_W$ then
the $\zeta$-instanton is well-approximated, away from the boundaries,
by a boosted soliton. There is some kind of ``emission amplitude''
and ``absorption amplitude'' associated with the region where the
boosted soliton joins the boundaries. In order to discuss these
we first consider the $\zeta$-instanton equation   on the
plane, but with some unusual boundary conditions at infinity.

\subsubsection{Fan Boundary Conditions}

\begin{figure}[h]
\centering
\includegraphics[scale=0.35,angle=0,trim=0 0 0 0]{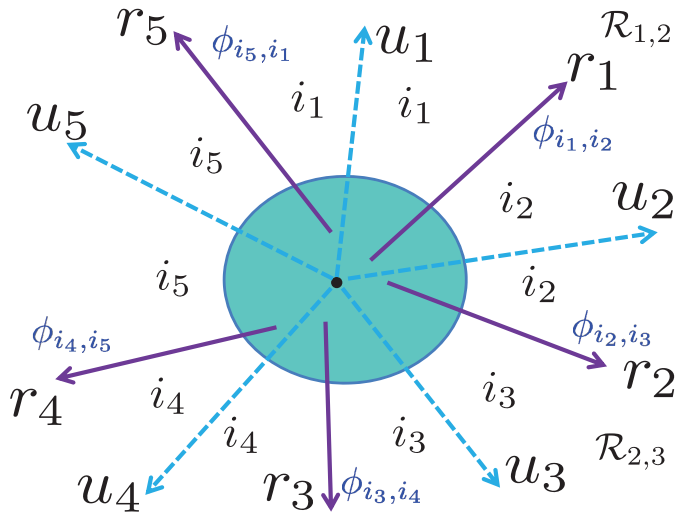}
\caption{\small Boundary conditions on the $\zeta$-instanton equation defined by
a cyclic fan of solitons. Here there are five boosted solitons and their worldlines near
infinity are the rays $r_1, \dots, r_5$. The boosted soliton solutions are exponentially
close to the constant vacua near the rays $u_1, \dots, u_5$ and can be modified there to
produce true solutions to the $\zeta$-instanton equation. }
\label{fig:WEDGES}
\end{figure}

We would like to consider solutions to the $\zeta$-instanton
equation that look like a collection of several boosted solitons at infinity.
The boosted solitons will be obtained from a cyclically ordered set of solitons
\be\label{eq:solseq}
\CF = \{ \phi_{i_1, i_2}^{p_1}, \dots, \phi_{i_n, i_1}^{p_n} \}.
\ee
ordered so that the worldlines of the boosted solitons have monotonically decreasing phase.
We refer to such a set of solitons as a \emph{cyclic fan of solitons}.
We are interested in solutions to the $\zeta$-instanton equation  which look like the corresponding
boosted solitons as $z$ moves clockwise around a circle at infinity,
as in Figure \ref{fig:WEDGES}.
Note this only makes sense when the phases of the successive differences
$z_{i_k, i_{k+1}}$ are clockwise ordered. We call such a sequence of
vacua a  \emph{cyclic fan of vacua}.

\begin{figure}[h]
\centering
\includegraphics[scale=0.35,angle=0,trim=0 0 0 0]{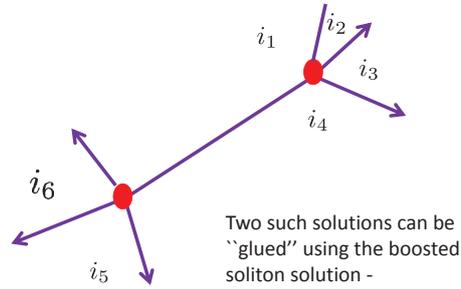}
\caption{\small Gluing two solutions with fan boundary conditions to produce a new
solution with fan boundary conditions. The red regions indicate where the
solution deviates significantly from the boosted solitons and the vacua.
When the ``centers'' of the two $\zeta$-instantons are far separated the
approximate, glued, field configuration can be corrected to a true solution.  }
\label{fig:GluedSolution}
\end{figure}

\begin{figure}[h]
\centering
\includegraphics[scale=0.35,angle=0,trim=0 0 0 0]{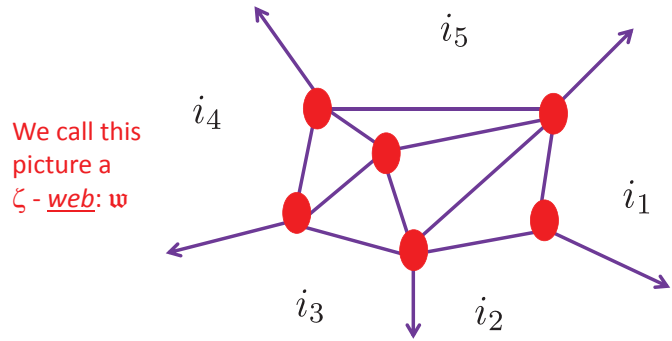}
\caption{\small Several solutions can be glued together to produce a $\zeta$-web solution  }
\label{fig:ZetaWeb}
\end{figure}

If the index of a certain Dirac operator is positive then
we expect, from index theory, that there will be $\zeta$-instantons which
approach such a cyclic fan of solitons at infinity. In fact, as mentioned
above, physicists studying domain wall junctions have   established the existence
of such solutions in some special cases \cite{Carroll:1999wr,Gibbons:1999np}
and there are even examples of exact solutions \cite{Oda:1999az}.
We will assume that a moduli space of such solutions $\CM(\CF)$ exists.
Based on physical intuition we expect these moduli spaces to satisfy two
crucial properties:

\begin{enumerate}

\item \emph{Gluing}: Under favorable conditions, two solutions which only differ significantly
from fan solutions inside a bounded region can be glued together as in Figure \ref{fig:GluedSolution}
This process can be iterated to produce what we call \emph{$\zeta$-webs},
shown in Figure \ref{fig:ZetaWeb}:

\item  \emph{Ends}: The moduli space $\CM(\CF)$ can have several connected components. Some of these
components will be noncompact, and the ``ends,'' or ``boundaries at infinity,''  of the moduli space will be described by
$\zeta$-webs.

\end{enumerate}


The compact connected components of $\CM(\CF)$-webs are called \emph{$\zeta$-vertices}.
We are most interested in the $\zeta$-vertices of dimension zero:
 These will contribute to the path integral of the LG model with fan
boundary conditions provided the fermion number of the outgoing states sums to $2$.
We claim that counting such points for fixed fans of solitons
produces interesting integers that   satisfy $L_\infty$ identities. We will state
that claim a bit more precisely later.
This picture is the inspiration for the web-formalism, to which we turn   next.
It will give us the language to state the above claim in more precise terms.

\subsection{The Web Formalism On The Plane}

We now switch to a mathematical formalism that we call the
\emph{web-based formalism} for describing the above physics.

\subsubsection{Planar Webs And Their Convolution Identity}

\bigskip
\noindent
\textbf{Definition}: The \emph{vacuum data} is
the pair $(\IV,z)$ where $\IV$ is a finite set
called the \emph{set of vacua} and
 $z: \IV \to \IC$ defines the \emph{vacuum weights}.
\bigskip

The vacua are denoted $i,j,\dots \in \IV$.
The vacuum weight associated to $i$ is denoted $z_i$.
  The vacuum weights $\{ z_i \}$ are
assumed to be in \underline{general position}. This means
\be\label{eq:VacWtSpace}
\{ z_1, \dots, z_N\} \in \CV:= \IC^N - \fE
\ee
where $\fE$ is the \emph{exceptional set}. The latter is
defined to be collections of vacuum weights that satisfy
at least one of the following three criteria:
(1)  $z_{ij} =0$ for some $i\not=j$ or (2)  three distinct vacuum weights are colinear
or (3) the weights allow the construction of an \emph{exceptional web}.
(Once we define webs below we can define exceptional webs to be those
whose deformation space has a dimension
larger than the expected dimension $2V-E$.)

\bigskip
\noindent
\textbf{Definition:} A \emph{plane web}
is a graph in $\IR^2$,   together with a
coloring of the \emph{faces}   by vacua
such that the labels across  each edge are different
and moreover, when oriented with
$i$ on the left and $j$ on the right the edge is
straight and parallel to the complex number
$z_{ij}:= z_i - z_j$. We take plane webs
to have all vertices of valence at least two.

\bigskip
\noindent
\textbf{Definition} The \emph{deformation type} of a
web is the equivalence class under stretching of
internal edges and overall translation. There is a
moduli space of deformation types and it can be
oriented. We denote an oriented deformation type
by $\fw$.

\begin{figure}[h]
\centering
\includegraphics[scale=0.28,angle=0,trim=0 0 0 0]{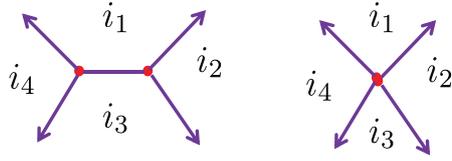}
\caption{\small The two webs shown here are considered to be   different
deformation types, even though the web on the left can clearly degenerate
to the web on the right.   }
\label{fig:DIFFERENT-DEFORMATION-TYPE}
\end{figure}

An example of two different deformation
types of web is shown in Figure \ref{fig:DIFFERENT-DEFORMATION-TYPE}.
In the web formalism and the related homotopical algebra there are
many tricky sign issues, ultimately tracing back to the
need to choose \underline{oriented} deformation types of webs.
Getting the signs right is a highly technical business and we will avoid it altogether
in these notes. That is not to say that the signs are unimportant - they most certainly
are! In \cite{Gaiotto:2015aoa} signs are taken into account with great care.

Next, we introduce some notation for certain fans of vacua
associated to a web. Recall that a fan of vacua is a   cyclically ordered set of vacua
so that successive edges are clockwise ordered. We can
associate two kinds of fans of vacua to a web $\fw$:

\begin{enumerate}

\item
\emph{The local fan of vacua at a vertex $v\in \fw$} is denoted $I_v(\fw)$.

\item
\emph{The fan of vacua at infinity} is denoted $I_\infty(\fw)$.

\end{enumerate}

\begin{figure}[h]
\centering
\includegraphics[scale=0.40,angle=0,trim=0 0 0 0]{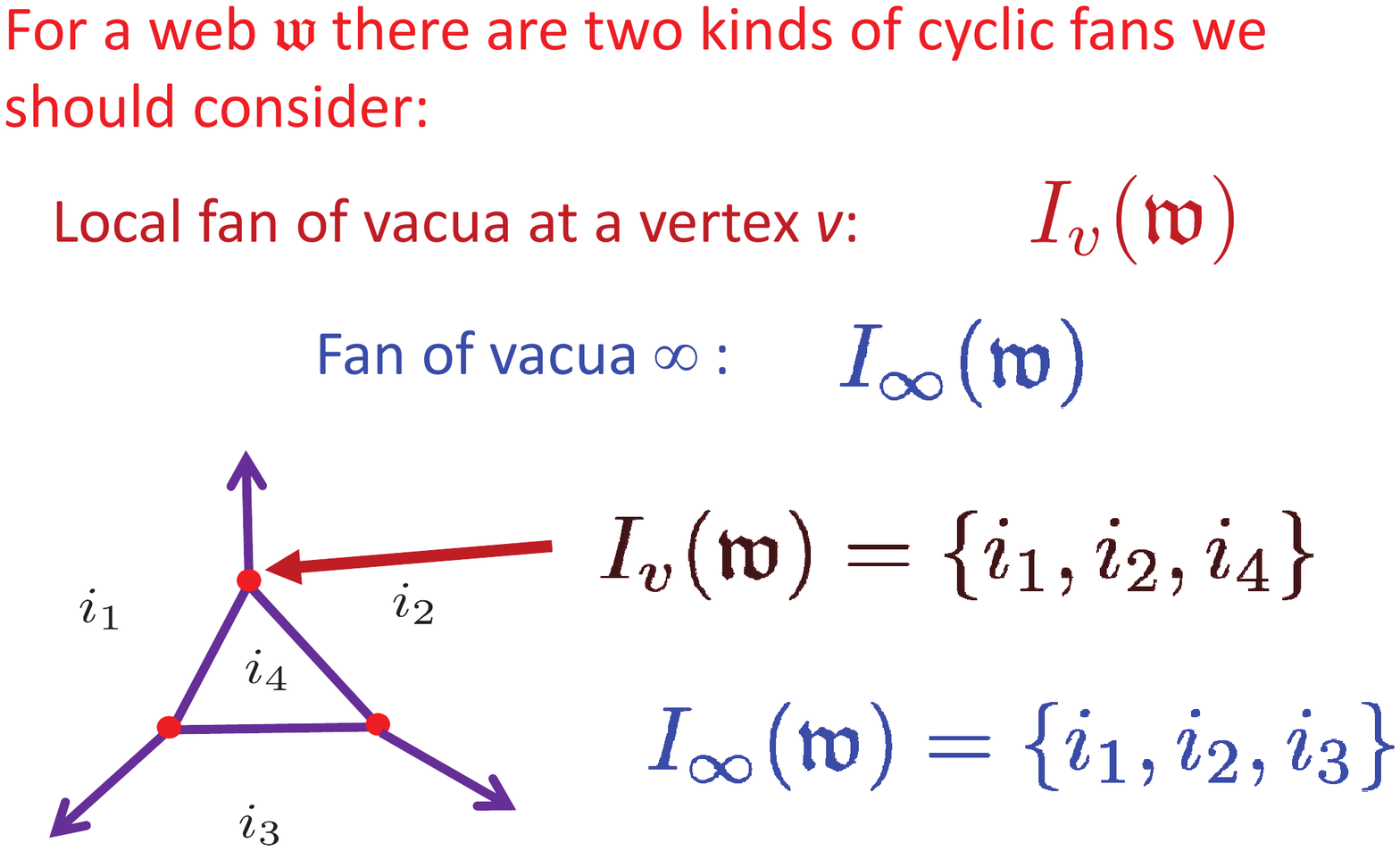}
\caption{\small Illustrating the local fan of vacua and the fan of vacua at infinity
for a web $\fw$.    }
\label{fig:LocalGlobalFan}
\end{figure}

See  Figure \ref{fig:LocalGlobalFan} for examples.

\begin{figure}[h]
\centering
\includegraphics[scale=0.35,angle=0,trim=0 0 0 0]{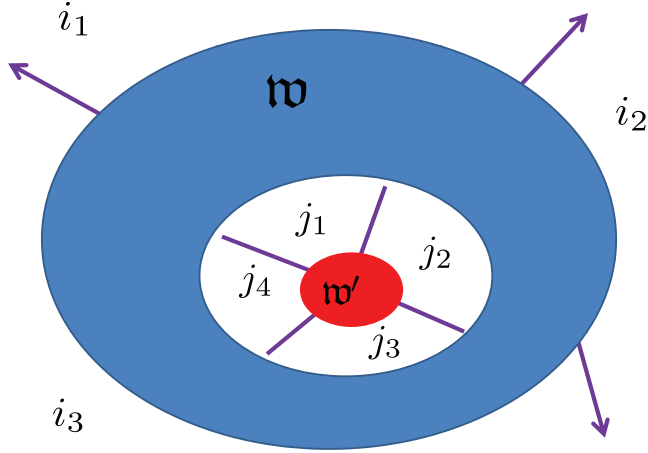}
\caption{\small Illustrating the convolution of a web $\fw$ with
internal vertex $v$ having a local fan $I_v(\fw)=\{j_1,j_2,j_3,j_4\}$ with
a web $\fw'$ having a fan at infinity $I_\infty(\fw') =\{j_1,j_2,j_3,j_4\}$.     }
\label{fig:CONVOLUTION}
\end{figure}

Now we introduce the key construction of a
\emph{convolution of webs}: Suppose we have two webs $\fw$ and $\fw'$ such that there is a vertex $v$ of $\fw$
where we have
 \be
 I_v(\fw) = I_\infty(\fw').
 \ee
Then define $\fw*_v \fw'$ to be the deformation type
 of a web obtained by cutting out a small disk around $v$
 and gluing in a suitably scaled and translated copy of
 the deformation type of $\fw'$.
 The procedure is illustrated in Figure \ref{fig:CONVOLUTION}.
 The upshot is that if  $\CW$ is the free abelian group generated by oriented deformation types of
webs then convolution defines a product
\be
\CW \times \CW \to \CW
\ee
(making it a ``pre-Lie algebra'' in the sense of \cite{ChapotonLivernet}).

\begin{figure}[h]
\centering
\includegraphics[scale=0.45,angle=0,trim=0 0 0 0]{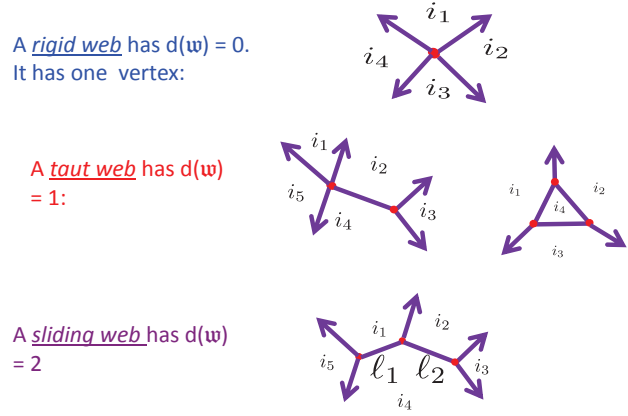}
\caption{\small Illustrating rigid, taut, and sliding webs with $0$, $1$, and $2$ internal
degrees of freedom. Here $d(\fw)$ refers to the dimension of the reduced moduli space of the
web, that is the dimension of the moduli space quotiented by the action of translation.   }
\label{fig:RigidTautSliding}
\end{figure}

Let us now consider the  \emph{taut webs}. These are, by definition,
those with only one internal degree of freedom. That is, the moduli
space of the taut webs is three-dimensional.  See Figure \ref{fig:RigidTautSliding}.
We define the \emph{taut element} $\ft\in \CW$  to be the sum over all the taut webs:
\be\label{eq:taut-planar}
\ft := \sum_{d(\fw)=3} \fw.
\ee
The key fact about taut webs is that
\be
\ft * \ft = 0.
\ee
The proof is that if we expand this out then we can group products in pairs which cancel.
The pairs correspond to opposite ends of a moduli space of ``sliding'' webs, with two
internal degrees of freedom. The idea is illustrated in Figure \ref{fig:TAUT-SQUARE}:

\begin{figure}[h]
\centering
\includegraphics[scale=0.45,angle=0,trim=0 0 0 0]{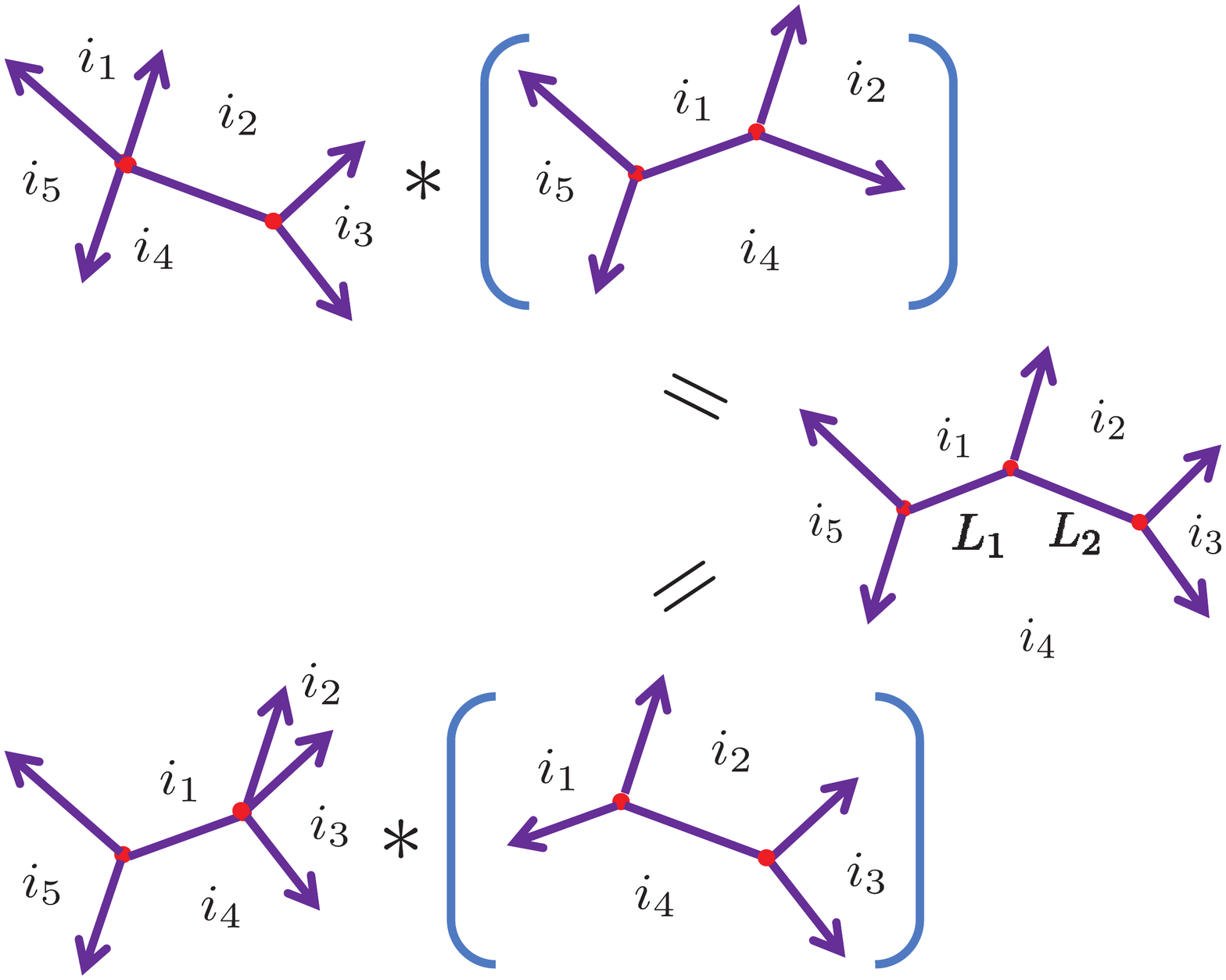}
\caption{\small The two boundaries of the deformation type of the sliding
web shown on the right correspond to different convolutions
shown above and below. If we use the lengths $L_1,L_2$
of the edges as coordinates then the orientation from the
top convolution is $dL_2 \wedge d L_1$. On the other
hand the orientation from the bottom convolution is
 $dL_1 \wedge d L_2$ and hence the sum of these two
 convolutions is zero. This is the key idea in the demonstration
 that $\ft*\ft=0$.   }
\label{fig:TAUT-SQUARE}
\end{figure}

\subsubsection{Representation Of Webs}

\textbf{Definition}:   A \emph{representation of webs} is
 a pair $\CR = ( \{ R_{ij} \}, \{ K_{ij} \} )$ where $R_{ij}$ are $\IZ$-graded $\IZ$-modules defined for
all ordered pairs $ij$ of distinct vacua and $K_{ij}$ is a degree $-1$ symmetric perfect  pairing
\be
K_{ij}: R_{ij} \otimes R_{ji} \to \IZ.\label{whichone}
\ee

Given a representation of webs, we define a representation of a  cyclic fan of vacua
$I = \{ i_1, i_2, \dots, i_n\}$ to be
\be
R_I := R_{i_1,i_2} \otimes R_{i_2,i_3} \otimes \cdots \otimes R_{i_n,i_1}
\ee
when $I$ is the cyclic fan at a vertex of a web we refer to $R_{I_v(\fw)}$ to
as the \emph{representation of the vertex}. Elements of $R_{I_v(\fw)}$
are called \emph{interior vectors}.

Next we collect the representations of all possible vertices by forming
\be\label{eq:Rint-def}
R^{\rm int} := \oplus_{I} R_I
\ee
where the sum is over all cyclic fans of vacua. We include $I=\emptyset$ and define
$R_{\emptyset}=\IZ$.  We want to define a map
\be
\rho(\fw): T \Rvtx  \to \Rvtx
\ee
where for any $\IZ$-module $M$ we define the tensor algebra to be
\be
TM := M \oplus M^{\otimes 2}  \oplus M^{\otimes 3}  \oplus \cdots
\ee
In fact, the operation will be graded-symmetric so it descends to a
map from the symmetric algebra $S\Rvtx \to \Rvtx$.

\begin{figure}[h]
\centering
\includegraphics[scale=0.25,angle=0,trim=0 0 0 0]{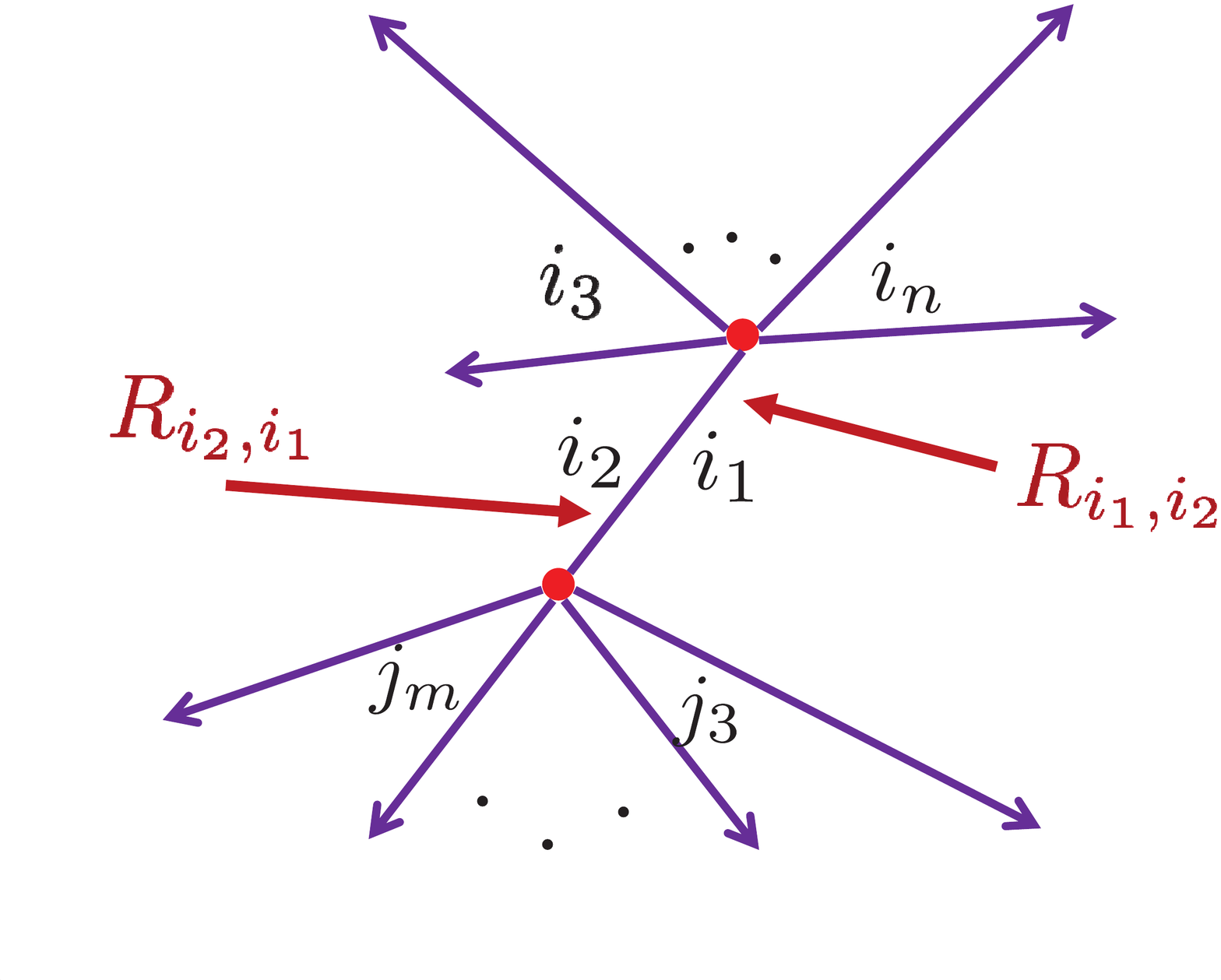}
\caption{\small The internal lines of a web naturally pair
spaces $R_{i_1,i_2}$ with $R_{i_2,i_1}$ in a web representation,
as shown here.    }
\label{fig:WEBEDGE}
\end{figure}

We now define the \emph{contraction operation}:
We take $\rho(\fw)[r_1, \dots, r_n]$ to be zero unless
$n = V(\fw)$, the number of vertices of $\fw$,
 and there exists an order $\{v_1, \dots , v_n \}$ for the vertices of $\fw$
such that $r_a \in R_{I_{v_a}(\fw)}$. If such an order exists, we will define our map
\be\label{eq:web-rep-1}
\rho(\fw): \otimes_{v \in \CV(\fw)}  R_{I_v(\fw)} \to R_{I_\infty(\fw)}
\ee
as the application of the contraction map $K$ to all internal edges of the web.
Here $\CV(\fw)$ is the set of vertices of $\fw$.
Indeed, if an edge joins two vertices $v_1, v_2 \in \CV(\fw)$ then if
$R_{I_{v_1}(\fw)}$ contains a tensor factor $R_{ij}$ it follows that
$R_{I_{v_2}(\fw)}$ contains a tensor factor $R_{ji}$ and these two
factors can be paired by $K$ as shown in Figure \ref{fig:WEBEDGE}.

It is not difficult to see that the convolution identity $\ft*\ft=0$ implies
 that $\rho(\ft)$ satisfies the axioms of an $L_\infty$ algebra $\rho(\ft): T\Rvtx \to \Rvtx$:
\be\label{eq:L-infty-rho}
\sum_{{\rm Sh}_2(S) } \epsilon_{S_1,S_2} ~ \rho(\ft) [ \rho(\ft)[S_1], S_2] = 0
\ee
where we sum over 2-shuffles of the ordered
set $S=\{ r_1, \dots, r_{n}  \}$ and $\epsilon_{S_1,S_2}$ is a sign factor
discussed at length in \cite{Gaiotto:2015aoa}.

\bigskip
\noindent
\textbf{Definition:} An \emph{interior amplitude} is an element $\beta \in \Rvtx$
of degree $+2$
so that if we define $e^\beta \in T \Rvtx \otimes \IQ$ by the exponential series
then
\be\label{eq:bulk-amp}
\rho(\ft)(e^\beta) = 0.
\ee

\bigskip
\noindent
\textbf{Definition:} A \emph{Theory} $\CT$ consists of a set of vacuum data  $(\IV,z)$,
a representation of webs $\CR=(\{ R_{ij}\}, \{ K_{ij}\} )$ and an interior amplitude $\beta$.
\bigskip

The simplest case of the $L_\infty$ equation implies there is a component of $\beta$ in
$R_{ij}\otimes R_{ji}$ sastifying a quadratic equation. Using $K$ we can interpret this component
of $\beta$
as a map $Q_{ij}: R_{ij} \to R_{ij}$ of degree one that squares to zero. Thus,
the $R_{ij}$ become chain complexes. It is also worth noting that if $\beta$ is an interior amplitude and we
 define $\rho_\beta(\fw)[r_1, \dots, r_\ell] := \rho(\fw)[r_1, \dots, r_\ell ,e^\beta]$
then   $\rho_{\beta}(\ft):T\Rvtx \to \Rvtx$ also satisfies the $L_\infty$ Maurer-Cartan
equation and in this way we obtain moduli spaces of Theories.

The mathematical structure we have just described is realized in the Landau-Ginzburg model as follows:
\begin{enumerate}

\item \emph{Vacua}:  $\IV$ is the set of critical points of $W$.

\item \emph{Vacuum weights}: $z_i = \zeta \bar W_i$

\item \emph{Web representation}:
\be\label{eq:Rij-WebRep}
R_{ij} := \oplus_{p \in \CS_{ij} } \IZ \Psi^{f+1}(p)
\ee
is the MSW complex, where we take the upper fermion number for each soliton $p$.
The contraction $K$ is defined by the path integral and is a kind of inner product
on the solitons.

\item \emph{Interior amplitude}:
Suitably interpreted, the path integral leads to a counting of
$\zeta$-instantons with fan boundary conditions and defines an
element in $R^{\rm int}$ which is an interior amplitude $\beta$.
This follows from localization of the path integral on the moduli
space of $\zeta$-instantons and the fact that the path integral
must create a $\CQ_{\zeta}$-closed state. For details
see Section \S 14 of  \cite{Gaiotto:2015aoa}.

\end{enumerate}

\subsubsection{Examples: Theories With Cyclic Weights  }\label{subsubsec:Examples}

Two useful examples have $\IV = \IZ/N\IZ$. We break the cyclic symmetry and
label vacua by $i\in \{ 0, \dots, N-1\}$ with weights:
\begin{equation}\label{eq:CyclicWt}
\IV^N_\vartheta: z_k = e^{- \I \vartheta - \frac{2 \pi \I}{N} k} \qquad k = 0, \cdots N-1
\end{equation}
(Although we have broken manifest cyclic symmetry all
physically relevant results are cyclically symmetric.
The web representations \eqref{eq:ExpleTN-webrep} and
\eqref{eq:SUN-Rij} below appear to violate this
symmetry but that is not the case when one takes into
account the   ``gauge freedom'' in the definition of fermion
numbers.)

\begin{figure}[h]
\centering
\includegraphics[scale=0.25,angle=0,trim=0 0 0 0]{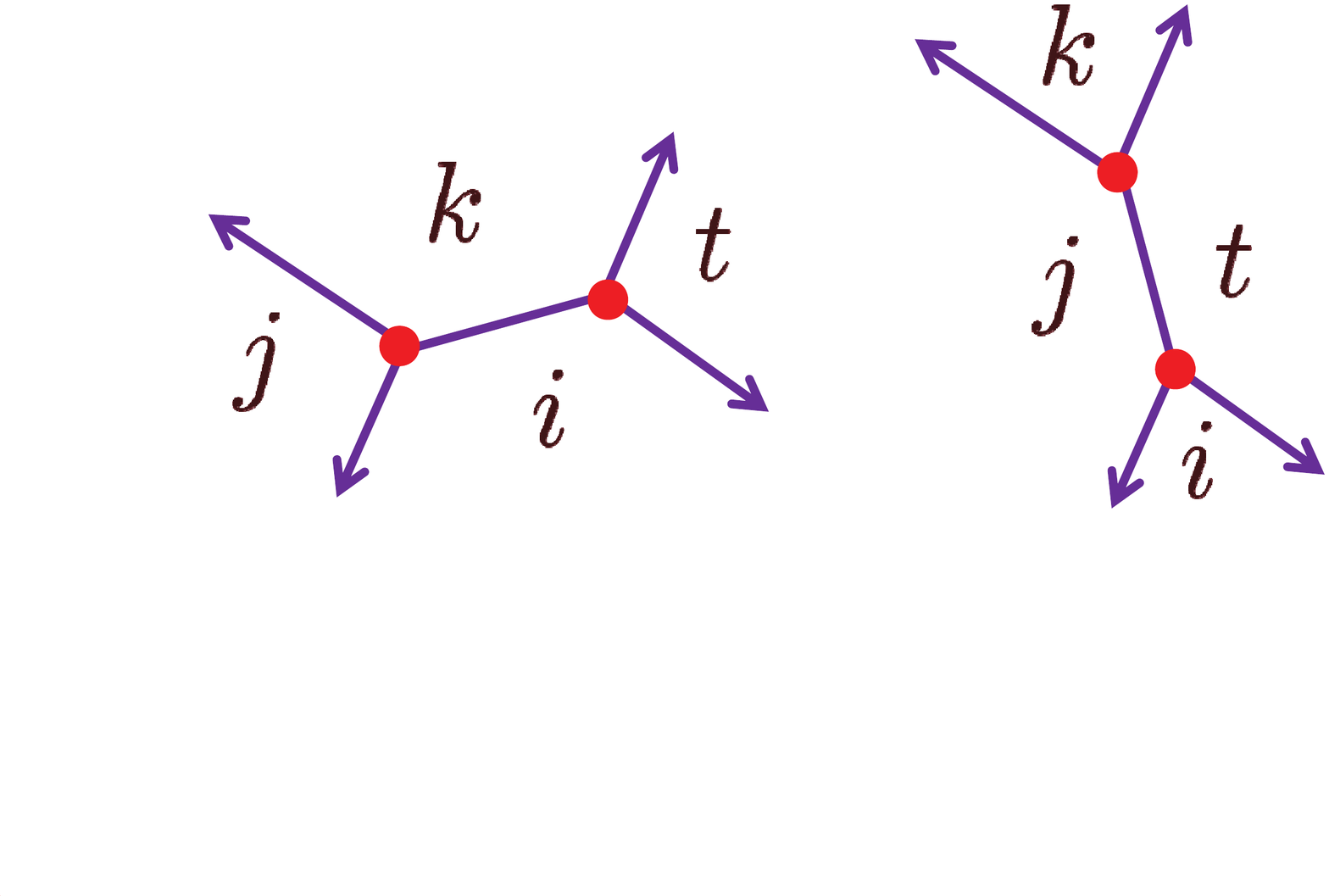}
\caption{\small The two terms in the component of the $L_\infty$ equations
for $i<j<k<t$. The resemblance to crossing symmetry is somewhat fortuitous. In other models
the $L_\infty$ equations do not resemble crossing symmetry equations.     }
\label{fig:TNEXAMPLE-1}
\end{figure}

The first example is   $\CT^N_{\vartheta}$  with a single chiral superfield
and superpotential
\be\label{eq:TN-Superpot}
W =     \phi - e^{-\I N \vartheta} \frac{\phi^{N+1}}{N+1}   .
\ee
The web-representation is
\footnote{The notation $\IZ[f]$ where $f$ is an integer means the following:
Recall that all modules in this paper are graded by $\IZ$ or a $\IZ$-torsor.  If $M$ is a graded module
then $M[f]$ denotes the module with grading shifted by $f$. When we write
$\IZ$ it is understood to have grading zero, so $\IZ[1]$ is the complex of rank
one concentrated in degree one.}
\begin{align}\label{eq:ExpleTN-webrep}
R_{ij} &= \IZ[1] \qquad i<j \cr
R_{ij} &= \IZ \qquad i>j.
\end{align}
At a vertex of valence $n$ we have $\deg R_I = n-1$ and hence only 3-valent
vertices contribute to the MC equations, so the only nonzero amplitudes
are $\beta_{ijk} \in R_{ijk}$ for $0 \leq i < j<k\leq N-1$. The $L_\infty$ equations
come from the two taut webs of Figure \ref{fig:TNEXAMPLE-1} and are just:
\be\label{eq:ExpleMC-1}
\beta_{ijk} \beta_{ikt} - \beta_{ijt} \beta_{jkt} = 0  \qquad  i<j<k<t
\ee

A more elaborate set of examples is provided by the mirror dual to the
B-model on $\IC \IP^{N-1}$ with $SU(N)$ symmetry. This again has
vacuum weights \eqref{eq:CyclicWt} but now we take

\begin{align}\label{eq:SUN-Rij}
R_{ij} &= A_{j-i}[1] \qquad i<j \cr
R_{ij} &= A_{N+j-i} \qquad i>j
\end{align}
where $A_\ell$ is the $\ell$-th antisymmetric power of a fundamental representation of $SU(N)$
and
\be\label{eq:Kij-TSUN}
K_{ij} ( v_1 \otimes v_2) = \kappa_{ij} \frac{ v_1 \wedge v_2}{\vol}
\ee
where $\kappa_{ij} $ is a sign (determined by a rule in \cite{Gaiotto:2015aoa})
and $\vol$ denotes a choice of volume form on $\IC^N$.  An $SU(N)$-invariant ansatz for the interior amplitude
reduces the $L_\infty$ MC equations to \eqref{eq:ExpleMC-1} above.

\subsection{The Web Formalism On The Half-Plane }

Fix a half-plane $\CH \subset \IR^2$ in the $(x,\tau)$ plane.
Most of our pictures will take the positive or negative half-plane, $x\geq x_\ell$ or $x\leq x_r$,
but it could be any half-plane.

\bigskip
\noindent
\textbf{Definition}:
Suppose $\p\CH$ is not parallel to any of the $z_{ij}$.
A  \emph{half-plane web } in $\CH$ is
a graph in the half-plane which may have some vertices (but no edges) on the boundary.
We apply the same rule as for plane webs: Label connected components
of the complement of the graph by vacua so that if the
 edges are oriented with $i$ on the left and
$j$ on the right then they  are parallel to $z_{ij}$. Boundary vertices
are allowed to be $0$-valent.

\begin{figure}[h]
\centering
\includegraphics[scale=0.3,angle=0,trim=0 0 0 0]{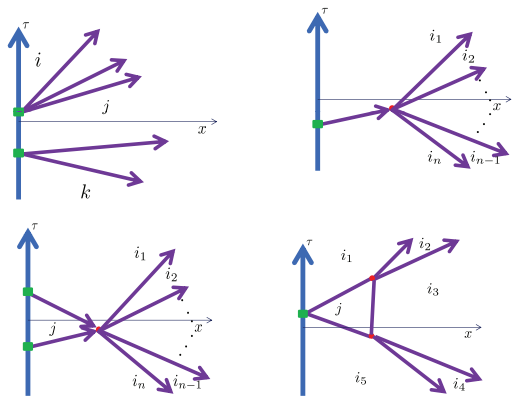}
\caption{\small Four examples of taut positive-half-plane webs   }
\label{fig:HALFPLANE-TAUTWEB}
\end{figure}

We can again speak of a deformation type of a half-plane web $\fu$.
Now translations parallel to the boundary of $\CH$ act freely on
the moduli space.
Once again we define half-plane webs to be \emph{rigid, taut,} and \emph{sliding}
if the reduced dimension of the moduli space is
$d(\fu) = 0,1,2 $, respectively. Similarly, we can define oriented deformation type
in an obvious way
and consider the free abelian group $\CW_{\CH}$ of oriented deformation types
of half-plane webs in the
  half-plane $\CH$. Some examples where $\CH= \CH_L$ is the positive half-plane  are shown in Figure \ref{fig:HALFPLANE-TAUTWEB}.

%
%
%

There are now two new kinds of convolutions:

1. Convolution at a boundary vertex defines
\be\label{eq:bbstar}
*: \CW_{\CH} \times \CW_{\CH} \to \CW_{\CH}
\ee

2. Convolution at an interior vertex defines:
\be\label{eq:bistar}
*: \CW_{\CH} \times  \CW \to \CW_{\CH}
\ee

\begin{figure}[h]
\centering
\includegraphics[scale=0.45,angle=0,trim=0 0 0 0]{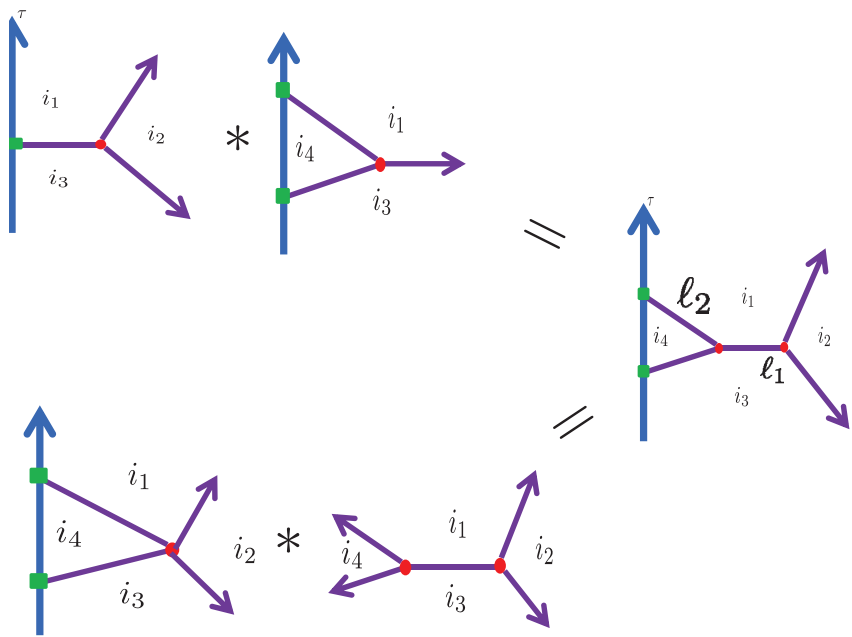}
\caption{\small An example of the identity on plane and half-plane
taut elements. On the right is a sliding half-plane web. Above is
a convolution of two taut half-plane webs with orientation
$dy \wedge d\ell_1 \wedge d\ell_2$, where $y$ is the vertical position of the
boundary vertex and $\ell_1, \ell_2$ are the lengths of the internal edges.  Below is a convolution
of a taut half-plane web with a taut plane web. The orientation is
$dy \wedge d\ell_2 \wedge d\ell_1$.  The two convolutions
determine the same deformation type but have opposite orientation, and
hence cancel. }
\label{fig:BLK-BDRY-WEBIDENT}
\end{figure}

We now define the half-space taut element:
\be
\ft_{\CH} := \sum_{d(\fu)=1} \fu.
\ee
The convolution identity is
\be
\ft_{\CH} * \ft_{\CH} + \ft_{\CH} * \ft_p =0.
\ee
where we now denote the planar taut element by $ \ft_p$.
The idea of the proof is the same as in the planar case. An example is
shown in Figure \ref{fig:BLK-BDRY-WEBIDENT}.

\subsection{Categorification Of The 2D Spectrum Generator }\label{subsec:Cat-Muij}

Given a half-plane and a representation of webs we can introduce
a collection of chain complexes $\widehat{R}_{ij}$ that will play an
important role in what follows.

One way to motivate the $\widehat{R}_{ij}$ is to recall the
Cecotti-Vafa-Kontsevich-Soibelman wall-crossing formula \cite{Cecotti:1992rm,Kontsevich:2008fj}
for the   Witten indices/BPS degeneracies $\mu_{ij}  = \Tr_{R_{ij}} (-1)^{{\rm \textbf{F}}}$ of 2d solitons.
The $\mu_{ij}$  were extensively studied in \cite{Fendley:1992dm,Cecotti:1992qh,Cecotti:1992rm}
where the wall-crossing phenomenon was first discussed. One way to state the wall-crossing formula
uses the matrix of BPS degeneracies
\be\label{eq:2d-CVKS-prod}
  \textbf{1}  + \oplus_{z_{ij}\in \CH} \widehat{\mu}_{ij} e_{ij} := \bigotimes_{z_{ij}\in \CH}
   (  \textbf{1} + \mu_{ij} e_{ij} )
\ee
where we assume there are $N$ vacua so we can identify
$\IV = \{1,\dots, N\}$,  $e_{ij}$ are  elementary $N\times N$ matrices,
  $\textbf{1}$ is the $N\times N$ unit matrix, and
 in the tensor product we order the factors left to right by the
 clockwise order of the phase of $z_{ij}$.
Continuous deformations of the K\"ahler metric $g_{I\bar J}$ and/or the superpotential $W$
in general lead to discontinuous changes in the number of solutions of
equations \eqref{eq:LG-flow}, \eqref{eq:left-infty-bc}, and \eqref{eq:right-infty-bc}. The deformations
of the K\"ahler metric do not change the indices $\mu_{ij}$ but changes in the superpotential that
cross walls where three or more vacuum weights become colinear  \underline{can} indeed change the BPS index $\mu_{ij}$.
The wall-crossing formula states that, nevertheless, the matrix \eqref{eq:2d-CVKS-prod} remains constant,
so long as no ray
through one of the $z_{ij}$ enters or leaves $\CH$.

The matrix \eqref{eq:2d-CVKS-prod} is sometimes called the ``spectrum generator.''
We now ``categorify'' the spectrum generator, and define $\widehat R_{ij}$ from  the formal product
\be\label{eq:Cat-KS-prod}
\SpecGen:= \oplus_{i,j=1}^N \widehat{R}_{ij} e_{ij} := \bigotimes_{z_{ij}\in \CH} (\IZ\cdot \textbf{1} + R_{ij} e_{ij} )
\ee
Note that $\widehat{R}_{ii}=\IZ$ is concentrated in degree zero
and $\widehat{R}_{ij} =0$ if $z_{ij}$ points in the opposite half-plane $-\CH$.
If $J= \{ j_1, \dots, j_n\} $ is a half-plane fan in $\CH$ then we define
\be \label{eq:RJ}
R_J:=  R_{j_1,j_2} \otimes \cdots \otimes R_{j_{n-1},j_n}
\ee
and $\widehat{R}_{ij}$ is just the direct sum over all $R_J$ for half-plane fans $J$ that begin with $i$
and end with $j$.

\bigskip
\noindent
\textbf{Remarks}:

\begin{enumerate}

\item We can ``enhance'' the (categorified) spectrum generator $\SpecGen$ with ``Chan-Paton factors.''
By definition,   \emph{Chan-Paton data} is  an assignment $i \to \CE_i$ of
a $\IZ$-graded module  to each vacuum $i\in \IV$. The modules $\CE_i$
will be   referred to as \emph{Chan-Paton factors}. The enhanced spectrum generator
is defined to be
\be\label{eq:Add-CP-Hop}
\SpecGen(\CE):=
\oplus_{i,j\in \IV} \widehat{R}_{ij}(\CE) e_{ij}:= \left( \oplus_{i\in \IV} \CE_i e_{ii} \right)
  \SpecGen  \left( \oplus_{j\in \IV} \CE_j e_{jj} \right)^*
\ee

\item  Phase ordered products such as \eqref{eq:2d-CVKS-prod} have also appeared
in many previous works on Stokes data, so the $R_{ij}$ can also be
considered to be ``categorified Stokes factors'' and $\SpecGen$ is an
``categorified Stokes matrix.''

\item  If we consider a family of theories where the rays $z_{ij}$ and $z_{jk}$
pass through each other then the categorified spectrum generator $\widehat{R}$
is in general \emph{not} invariant, in striking contrast to \eqref{eq:2d-CVKS-prod}.
A categorified version of the Cecotti-Vafa-Kontsevich-Soibelman wall crossing formula
is a rule for describing how $\widehat{R}$ changes. We will discuss such a rule in
 \S \ref{subsec:Cat-CVWC} below.

\end{enumerate}

\subsection{\afty-Categories Of Thimbles And Branes}

\subsubsection{The \afty-Category Of Thimbles}

We now want to define the   \afty-\emph{category of Thimbles}, denoted $\fVac$:  Suppose we are given the data of a
Theory $\CT$ and a half-plane $\CH$. Then $\fVac$  has as objects
the vacua $i,j,\dots \in \IV$. As  we will see,the objects of the category are better thought of
as  Thimble branes $\fT_i,\fT_j,\dots$, defined at the end of \S \ref{subsec:VacCategory} below.
The space of  morphisms $\Hom(j,i)$ is simply
\be
\Hom(j,i):=\Hop(i,j) := \widehat{R}_{ij}.
\ee
Here we have also introduced the notation    $\Hop(i,j):=\Hom(j,i)$ since many formulae in \afty-theory
look much nicer when written in terms of $\Hop$.

We can enhance the category with Chan-Paton factors.
The morphism spaces are simply the matrix elements of $\widehat{R}(\CE)$:
\be
\Hop^\CE(i,j):=\widehat{R}_{ij}(\CE) = \CE_i \widehat{R}_{ij} \CE_j^*.
\ee
The corresponding category is denoted $\fVac(\CE)$.

Now we need to define the \afty-multiplication in $\fVac(\CE)$
of an $n$-tuple of composable morphisms.
As a first step, for any half-plane web $\fu$ we define   a map
\be
\rho(\fu): T\SpecGen(\CE) \otimes T\Rvtx \to  \SpecGen(\CE)
\ee
It will be graded symmetric on the second tensor factor.
As usual, we define the element
\be\label{eq:Rho-rp-r}
\rho(\fu)[r^\p_1, \dots, r^\p_m;r_1, \dots, r_n]
\ee
by contraction.
We will abbreviate this to $\rho(\fu)[P;S] $ where
 $P= \{r^\p_1, \dots, r^\p_m\}$ and $S= \{ r_1, \dots, r_n \}$.
We define $\rho(\fu)[P;S] $  to be zero  unless the following conditions hold:
\begin{itemize}
%
%

\item The boundary arguments match in order and type those of the boundary vertices: $r^\p_a \in R_{J_{v^\p_a}(\fu)}(\CE)$.
\item We can find an order of the interior vertices $\CV_i(\fu)=\{v_1, \dots, v_n \}$ of $\fu$ such that they match the order and type of the interior arguments: $r_a \in R_{I_{v_a}(\fu)}$.
\end{itemize}

If the above
 conditions hold, we will simply contract all internal lines with $K$ and contract the Chan Paton elements of consecutive pairs of
$r^\p_a$ by the natural pairing $\CE_{i} \otimes \CE_j^* \to \delta_{ij} \IZ$.
With this definition in hand, we can check that the
convolution identity for taut elements implies
a corresponding identity for $\rho[\ft_\CH]$:
\be\label{eq:big-rel-rho}
  \sum_{{\rm Sh}_2(S), {\rm Pa}_3(P)} \epsilon ~  \rho(\ft_{\CH})[P_1,
\rho(\ft_{\CH})[P_2 ; S_1 ], P_3; S_2] \\
 +
\sum_{{\rm Sh}_2(S)}   \epsilon ~  \rho(\ft_{\CH})[ P;
\rho(\ft_p)[  S_1 ],  S_2] = 0  .
\ee
where ${\rm Pa}_3(P)$ is the set of partitions of the ordered set $P$ into an ordered set of
three disjoint ordered sets, all inheriting the ordering of $P$. The signs are discussed
in detail in \cite{Gaiotto:2015aoa}. We call
\eqref{eq:big-rel-rho} the $LA_\infty$ relations.

\begin{figure}[h]
\centering
\includegraphics[scale=0.35,angle=0,trim=0 0 0 0]{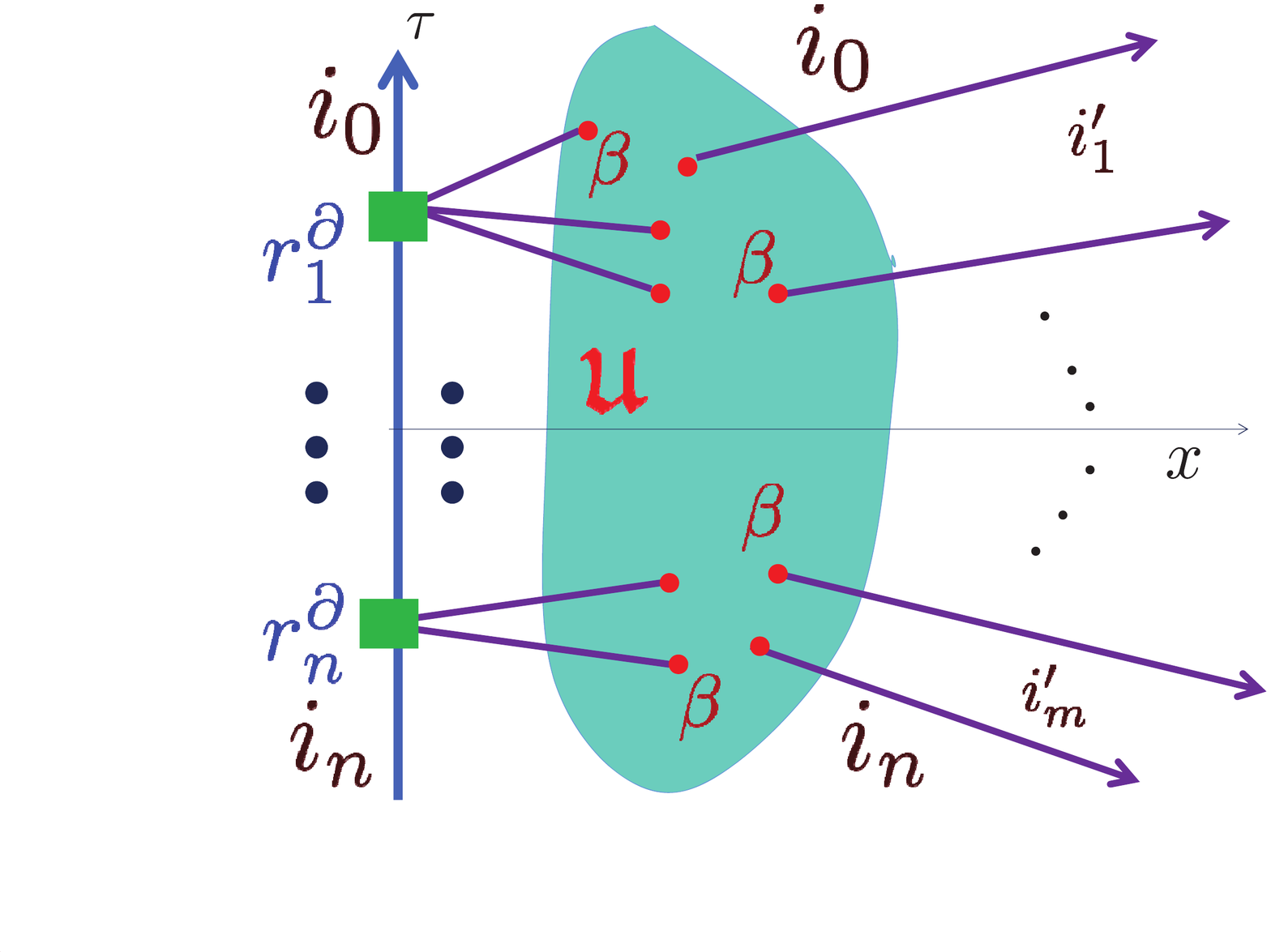}
\caption{\small Illustrating the \afty-product on time-ordered boundary vectors
$r^\p_1, \dots, r^\p_n$. We sum over taut half-plane webs $\fu$, indicated
by the green blob, and saturate all interior vertices with the interior
amplitude $\beta$.
   }
\label{fig:AINFTY-PRODUCT}
\end{figure}

The most important consequence of these identities is that if we are given an interior amplitude $\beta$,
we can immediately produce an $A_\infty$ category where the multiplication
\be\label{eq:R-AFTYALG}
\rho_\beta(\ft_\CH): T\SpecGen(\CE) \to \SpecGen(\CE)
\ee
is defined by saturating all the interior vertices with the interior amplitude:
\be
 \rho_\beta(\ft_\CH)[r^\p_1, \dots, r^\p_m] := \rho(\ft_\CH)[r^\p_1, \dots, r^\p_m;e^\beta].
\ee
This has the effect of killing the second term in \eqref{eq:big-rel-rho} and combining the
first summand into the usual defining relations for an \afty-category.
The product is illustrated in Figure \ref{fig:AINFTY-PRODUCT}.

\bigskip
\noindent
\textbf{Remark}: The conceptual meaning of \eqref{eq:big-rel-rho} is that
there is an $L_\infty$ morphism from the $L_\infty$ algebra $\Rvtx$ to
the $L_\infty$ algebra of the Hochschild cochain complex of the $A_\infty$
category $\fVac(\CE)$. The paper \cite{Kapranov:2014uwa} shows that in the
present context the map is in fact an $L_\infty$ isomorphism.

\subsubsection{The   \afty-Category  Of Branes }\label{subsec:VacCategory}

We define a \emph{Brane}, denoted $\fB=(\CE,\CB)$ to be a
choice of Chan-Paton data $\CE$ together with a
\emph{boundary amplitude}, that is, a degree $+1$ element
\be
\CB \in \SpecGen(\CE)
\ee
that solves the Maurer-Cartan equations
\be\label{eq:boundary-amp}
\sum_{n=1}^\infty \rho_\beta(\ft_\CH)[\CB^{\otimes n}] = \rho_\beta(\ft_\CH)[\frac{\CB}{1-\CB}] = 0.
\ee

The category of  Branes is denoted $\fB\fr$. It depends on the Theory $\CT$ and
the half-plane $\CH$. Its objects are Branes $\fB=(\CE,\CB)$ where $\CE$  is  \emph{any} choice of
 Chan Paton data $\CE$ and $\CB$ is a  compatible boundary amplitude.
The space of morphisms from $\fB_2$  to $\fB_1$ is defined by simply modifying the
enhanced spectrum generator to
\be\label{eq:BHOM}
\Hop(\fB_1, \fB_2) := \left( \oplus_{i} \CE_i^1 e_{ii} \right)\otimes \SpecGen  \otimes \left( \oplus_{i} \CE_i^2 e_{ii} \right)^*.
\ee
In order to  define the composition of morphisms
\be
\delta_1 \in \Hop(\fB_0, \fB_1), \quad \delta_2 \in \Hop(\fB_1, \fB_2) ,  \dots, \delta_n \in
\Hop(\fB_{n-1}, \fB_n)
\ee
we use  the formula
\be\label{eq:BraneMultiplications}
M_n(\delta_1,\dots, \delta_n) :=
   \rho_\beta(\ft_{\CH})\left(\frac{1}{1-\CB_0},\delta_1, \frac{1}{1-\CB_1}, \delta_2, \dots, \delta_n, \frac{1}{1-\CB_n} \right).
\ee
Note that $M_n(\delta_1,\dots, \delta_n)\in \Hop(\fB_0,\fB_n)$.
After some work (making repeated use of the fact that the $\CB_a$ solve the \afty-Maurer-Cartan equation)
one can show that the $M_n$ satisfy the \afty-relations and hence $\fB\fr$ is an \afty-category. In
particular $M_1^2=0$ can be considered to be a differential (i.e. a nilpotent supercharge).

\bigskip
\noindent
\textbf{Remarks}:

\begin{enumerate}

\item The multiplication \eqref{eq:BraneMultiplications} can be illustrated much
as in Figure \ref{fig:AINFTY-PRODUCT}. The only difference is that now the boundary
vectors $r_s^\p$ don't have to saturate all boundary vertices. Rather, boundary vertices
between $r_k^\p$ and $r_{k+1}^\p$ can be saturated by the boundary amplitude $\CB_k$.

\item For each vacuum $i$ we define the Thimble Brane $\fT_i$ to be the brane
with CP data $\CE(\fT_i)_j = \delta_{i,j} \IZ$ with boundary amplitude $\CB(\fT_i)=0$.
Then the category of Thimbles $\fVac$ is a full subcategory of $\fB\fr$.

\end{enumerate}

\subsubsection{Realization In The LG Model }\label{subsubsec:BraneAmpLG}

\begin{figure}[h]
\centering
\includegraphics[scale=0.35,angle=0,trim=0 0 0 0]{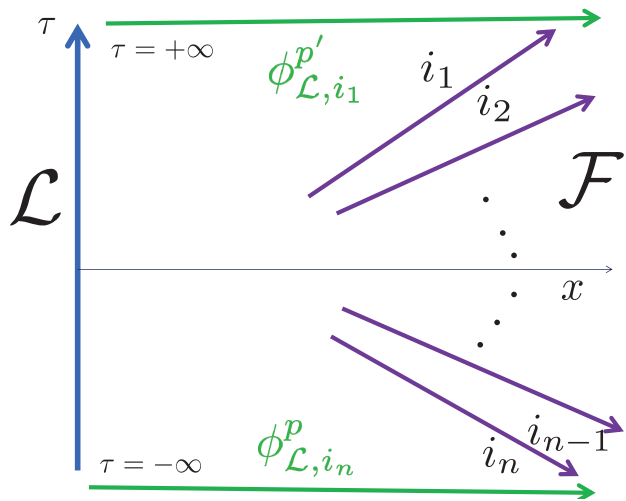}
\caption{\small Boundary conditions for general half-plane instantons with
fan boundary conditions at $x\to + \infty$ and solitons at $\tau \to \pm \infty$.
   }
\label{fig:HALFPLANEBC}
\end{figure}

Choose $\CH$ to be the positive half-plane with boundary conditions
set by a Lagrangian $\CL\subset X$. The Chan-Paton data is given
by the MSW complex:
\be
\CE_i = \IM_{\CL, i}
\ee
We consider amplitudes with boundary conditions shown in
Figure \ref{fig:HALFPLANEBC}. The counting of the number of $\zeta$-instantons
satisfying these boundary conditions can be used to
define an element $\CB_J \in \CE\otimes R_J \otimes \CE^*$.
As with the case of the interior amplitude, localization of the path integral to
the moduli space of $\zeta$-instantons together with $\CQ_\zeta$-closure of the
state produced by the path integral implies that $\CB$ is a boundary amplitude
in the above sense.

In general the $M_1$-cohomology of  $\Hop(\fB_1, \fB_2)$ is a
space of $\CQ_\zeta$-closed local boundary operators and the physical
interpretation of $M_n(\delta_1, \dots, \delta_n)$ is that we are taking
a kind of ``operator product.''  The $\CQ_{\zeta}$ closure of the path integral
implies that the $M_n$ satisfy the \afty-MC equation.

\textbf{Remarks}:

\begin{enumerate}

\item Spaces of local operators between some simple branes, such as
Thimbles, for the theories with cyclic weights (Section \S \ref{subsubsec:Examples} above)
are described in Section \S 5.7 of \cite{Gaiotto:2015aoa}. In the theory with
$SU(N)$ symmetry they are nontrivial representations of $SU(N)$.

\item If we want good morphism spaces associated to the interval $[x_\ell, x_r]$
we need to restrict the class of Lagrangian submanifolds, as we have seen.
In \cite{Gaiotto:2015aoa}
it is argued that the suitable class of branes for which the web-formalism makes
sense is the class of  \emph{$W$-dominated branes} for
which $\Im(\zeta^{-1} W) \to +\infty$ at infinity. (For right-branes on boundaries of the
negative half-plane we require $\Im(\zeta^{-1} W) \to -\infty$.) This class of branes
includes the union of branes of class $T_{\kappa}$ for $\kappa$ in the open half-plane
containing $\zeta$. However, in order to compare to the Fukaya-Seidel category one should restrict to a smaller
class of branes, and it turns out that the subset of branes of class $T_{\zeta}$ will suffice.
This might seem odd, since, as mentioned above, in our formulation of the
Fukaya-Seidel category, we definitely want to
use branes of type $T_{\kappa}$ with $\kappa \not=\pm \zeta$.
The reason for the apparent discrepancy is explained in the next section.

\end{enumerate}

\subsection{Relation Of The Web-Based Formalism To The FS Category }

Now we would like the relate the \afty-category constructed in the FS
approach and in the web-based approach, say, for the positive half-plane.
The  web-based formalism applies to   branes of class $T_\zeta$ and
our description of the FS  category applies to branes of class $T_\kappa$ with $\kappa \not= \pm \zeta$.

To relate the two we strongly use the rotational \underline{non}-invariance
of the $\zeta$-instanton equation and  consider the FS category based on branes of class $T_\zeta$ but
now the morphism spaces are defined by solving the equation on a \underline{horizontal} strip, obtained
from the vertical one by rotation by  $\pi/2$. Thus, to define the morphisms of the FS category we use the
 MSW complex  $\IM_{\fB_1, \fB_2}$ whose  generators are     solutions of the $\zeta$-instanton
 equation are invariant under translation in $x$ (but not in $\tau$). Now we can use
branes of class $T_\zeta$ on the upper and lower boundary.

\begin{figure}[h]
\centering
\includegraphics[scale=0.3,angle=0,trim=0 0 0 0]{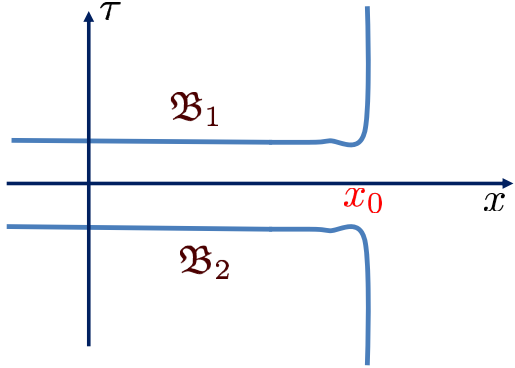}
\caption{\small We count rigid $\zeta$-instantons in the funnel geometry
to define an \afty-morphism between the FS category and the web-based category. The
branes $\fB_1, \fB_2$ are in class $T_\zeta$.
   }
\label{fig:FUNNEL-STRIP}
\end{figure}

To relate the FS and web-based categories we now consider the $\zeta$-instanton equation
on the funnel geometry of Figure \ref{fig:FUNNEL-STRIP}.
A state in the far past at $x\to - \infty$ on the strip is an incoming soliton, in the above sense.
A state in the morphisms in the web-based
formalism gives half-plane fan boundary conditions at infinity for the positive
half-plane. But these two states determine boundary conditions for the
 $\zeta$-instanton equation on the space in Figure \ref{fig:FUNNEL-STRIP}. We can therefore define
a map
\be
\CU: \IM_{\fB_1,\fB_2} \to \Hop(\fB_1, \fB_2)
\ee
The matrix elements of $\CU$ are defined by counting $\zeta$-instantons in the funnel geometry.
When we consider states of the same fermion number the expected dimension of the
moduli space is dimension zero and the moduli space is expected to be a finite set of points.

\begin{figure}[h]
\centering
\includegraphics[scale=0.45,angle=0,trim=0 0 0 0]{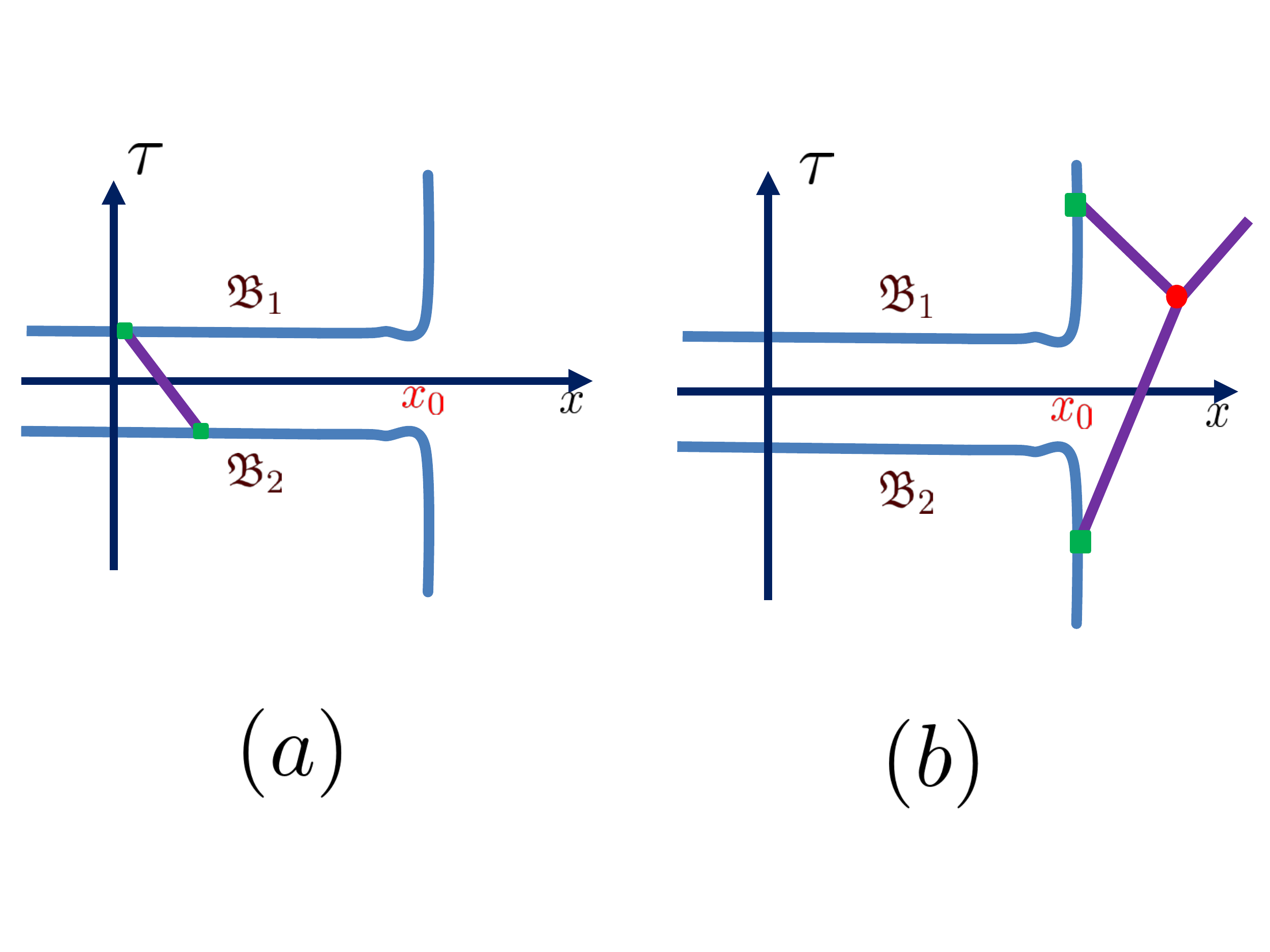}
\caption{\small When the difference of fermion numbers of ingoing
and outgoing states is $+1$ there will be a one-dimensional moduli space
of $\zeta$-instantons. The two typical boundaries are indicated in (a)
and (b). They lead to the two terms in the equation assuring that
$\CU$ is a chain map.
   }
\label{fig:FUNNEL-CHAINMAP}
\end{figure}

We claim that $\CU$ is a chain map. To prove this we consider the one-dimensional moduli spaces
of solutions to the $\zeta$-instanton equation between states whose fermion number
differs by $1$. The two ends correspond to $\zeta$-instantons far down the strip -
giving the differential on $\IM_{\fB_1,\fB_2} $   and taut webs far out on the
positive half-plane, giving the differential on $\Hop(\fB_1,\fB_2)$, so
\be
\CU \circ M_1^{\rm FS} -  M_1^{\rm web}\circ \CU  =0
\ee
where $M_1$ denotes the differential on the morphisms in the \afty-category. Using similar arguments
one can show that  $\CU$ can
be extended to a full   \afty-equivalence between the categories. For further details
see Section \S 15 of \cite{Gaiotto:2015aoa}.

\subsection{Local Operators}

The web formalism can also be used to determine spaces of local operators. To do
this, we extend the $L_\infty$ algebra $R^{\rm int}$ by introducing a module $R_i \cong \IZ$
for each vacuum $i\in \IV$. In Section \S  9  of \cite{Gaiotto:2015aoa} we show
that
\be\label{eq:Rc}
R_c: = \oplus_{i\in \IV} R_i  \oplus R^{\rm int}
\ee
admits a natural $L_\infty$ algebra structure associated with \emph{doubly-extended webs}.
The extra data we add to a web are vertices with no edges attached. We argue that the
cohomology of this complex is a space of local operators. The realization of these
local operators in the Landau-Ginzburg models is a little subtle and is discussed
in detail in   Section \S 16 of \cite{Gaiotto:2015aoa}. The $R_i$ have
generators corresponding to  an insertion of ``closed string'' states on the circle with $\phi(x) = \phi_i$,
while the $R_I\subset R^{\rm int}$ are related to \emph{twisted} $\zeta$-solitons.
That is, solitons on the circle where $\zeta(x) = \zeta e^{\I x}$.  It turns out that
the local operators described by the $M_1$-cohomology of $\Hop(\fB_1,\fB_2)$ and the
cohomology of $R_c$  include certain kinds of disorder operators, novel to Landau-Ginzburg
theories.

As an example, including suitable disorder operators helps resolve a puzzle in mirror
symmetry: The standard $B$-model local operators of the $\IC\IP^{N-1}$ model do not
correspond to the standard $A$-model local operators of the affine $SU(N)$ Toda model.
Nevertheless, as shown in \cite{Gaiotto:2015aoa} the cohomology of \eqref{eq:Rc}
beautifully reproduces the space of $B$-model operators on $\IC\IP^{N-1}$.

\section{Interfaces And Categorified Wall-Crossing}\label{sec:Interface-WC}

\subsection{Motivation: Interfaces In Landau-Ginzburg Models}\label{subsec:LG-Interface}

Suppose we have a family  of superpotentials $W(\phi;c)$, parametrized
by a point $c$ in a topological space $C$.
\footnote{$C$ can be any space, but the notation is again chosen because
one of the primary motivations is the theory of spectral networks and Hitchin systems.}
Suppose $\wp: [x_\ell, x_r] \to C$ is a continuous path.
Then we can define a variant of LG theory based on
 an $x$-dependent superpotential:
\be
W_x(\phi):=  W(\phi;\wp(x)),
\ee
 so that
$W_x(\phi)$ is constant (in $x$)  for $x\leq x_\ell$ and for $x\geq x_r$.
Clearly this $1+1$ dimensional theory no longer has translational
invariance. It does, however, still have two out of the four
supersymmetries of LG theory. This is demonstrated most easily if
we take   the approach via  Morse theory/SQM using the Morse
function on $\Map(\IR,X)$:
\be\label{eq:Interface-h}
h = -   \int_{\IR}
\left[\phi^*(\lambda) - \half {\Re}(\zeta^{-1} W(\phi;\wp(x) )dx \right].
\ee
Clearly the resulting theory has a kind of ``defect'' or
``domain wall'' localized near $[x_\ell, x_r]$ interpolating
between the left LG theory defined with superpotential $W_{x_{\ell}}(\phi)$
and the right LG theory defined with  superpotential $W_{x_r}(\phi)$.
We will refer to this as a (LG, supersymmetric) \emph{interface}.
The term ``Janus'' is also often used in the literature.

\begin{figure}[h]
\centering
\includegraphics[scale=0.35,angle=0,trim=0 0 0 0]{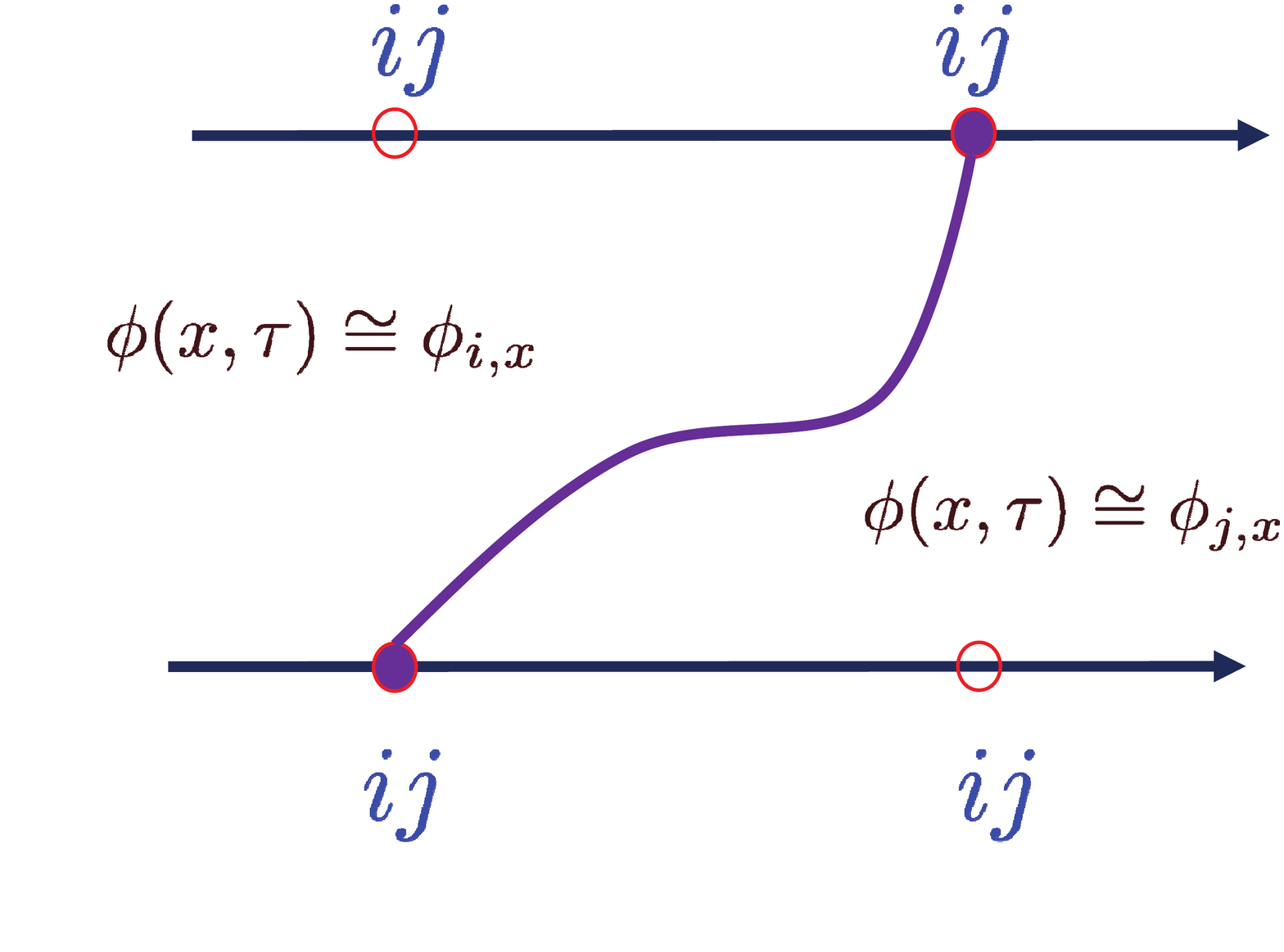}
\caption{\small An analog of the boosted soliton for the case of a supersymmetric
interface.   }
\label{fig:FORCEDFLOWBOOSTEDSOLITON}
\end{figure}

In the above setup we have a continuous family of vacuum weights
\be
z_i(x) = \zeta \bar W_x(\phi_{i,x})
\ee
where the vacuum $i$ is parallel transported from the vacua in the
theory at $x_{\ell}$ and $\phi_{i,x}$ are the critical points
of the superpotential $W_x(\phi)$. The $\zeta$-instanton equation now becomes:
\be\label{eq:LG-forced-flow}
 \left( \frac{\p}{\p x} + \I \frac{\p}{\p \tau}\right) \phi^I = \frac{\I \zeta}{2} g^{I \bar J}
\frac{\p \bar W}{\p \bar \phi^{\bar J}}(\bar\phi;\wp(x))
\ee
and $\zeta$-solitons are just $\tau$-independent solutions. The  analog of boosted solitons have
curved worldlines, as in Figure \ref{fig:FORCEDFLOWBOOSTEDSOLITON}

Now, we would like to define a relation of the branes in the left theory to the
branes in the right theory by ``parallel-transporting'' across the interface.

\subsection{Abstract Formulation: Flat Parallel Transport Of Brane Categories}

Suppose we have a ``continuous family of Theories.'' We use the term
``Theory'' in the sense of the web formalism. To make sense of this
one must put a topology on the set of Theories. Note that the set of
vacuum weights $\CV$ of \eqref{eq:VacWtSpace} carries a natural topology.
Thus we can certainly speak of a continuous map
\be\label{eq:tame}
\wp: [x_{\ell}, x_r] \to \CV = \IC^N - \fE
\ee
We call this a  \emph{vacuum homotopy}.

More generally, one can also define a sense in which web representations and
the interior amplitudes change continuously. So, in general, we have a continuous family of Theories $\CT(x)$
on $[x_\ell, x_r]$. We would like to relate $\CT^\ell = \CT(x_\ell)$ to $\CT^r = \CT(x_r)$.
More precisely, we want to define an \afty-functor
\be
\CF(\wp): \fB\fr(\CT^\ell, \CH ) \rightarrow \fB\fr(\CT^r,\CH)
\ee
where $\CH$ is, say, the positive half-plane.

The functor $\CF(\wp)$ is meant to be a categorical version of  parallel transport
by a flat connection.
Thus we want:

\begin{enumerate}

\item An \afty-equivalence of functors:
\be
\CF(\wp_1) \circ \CF(\wp_2) \cong \CF(\wp_1 \circ \wp_2)
\ee
for composable paths $\wp_1, \wp_2$.

\item An \afty-equivalence of functors:
\be
\CF(\wp_1)  \cong \CF( \wp_2)
\ee
for  paths $\wp_1, \wp_2$ homotopic in some suitable space.

\end{enumerate}

We will show that one \underline{can} construct such functors for
``tame'' vacuum homotopies, i.e. homotopies of the type \eqref{eq:tame}. Flushed with
success we then \underline{want} to extend the construction to more general
vacuum homotopies for paths of weights which cross the exceptional walls $\fE$.
But you don't always get what you want:

\bigskip
\bigskip
\fbox{\fbox{\parbox{5.5in}{The   existence of such a
functor \underline{forces}  discontinuous changes of the web representation
and the interior amplitude:
This is the categorified version of wall-crossing.
}}}
\bigskip

The secret to constructing $\CF(\wp)$ is the theory of \emph{Interfaces}
in the web-based formalism, to which we turn next.

\subsection{Interface Webs And Composite Webs}

\subsubsection{The \afty-Category Of Interfaces}

In order to understand the parallel transport of Brane
categories it will actually be very useful to consider
\emph{discontinuous} jumps between Theories.

Given a pair of vacuum data $(\IV^-, z^-)$ and $(\IV^+,z^+)$
we can define an interface web by using the data on the
negative and positive half-planes, respectively.
Examples are shown in Figures  \ref{fig:DOMAINWALL-CHANPATON}
and \ref{fig:ID-INTERFACE}  below. We can define the taut
element $\ft^{-,+}$ and write a convolution identity.
Next, if   we are given left and right Theories $(\CT^-, \CT^+)$ then we can define
a representation of interface webs:

\begin{enumerate}

\item  Chan-Paton factors are now labeled by  a \underline{pair} of vacua $\CE_{j_-, j'_+}$.

\item
At a boundary vertex we have the representation:
\be \label{eq:RJ-intfc}
R_J(\CE):= \CE_{j_m,j'_1} \otimes R_{J'_+}^+
 \otimes \CE_{j_1,j'_n}^* \otimes R_{J_-}^-  .
\ee
associated to the picture in Figure \ref{fig:DOMAINWALL-CHANPATON},
where $J=(J_-, J_+')$.

\end{enumerate}

\begin{figure}[h]
\centering
\includegraphics[scale=0.35,angle=0,trim=0 0 0 0]{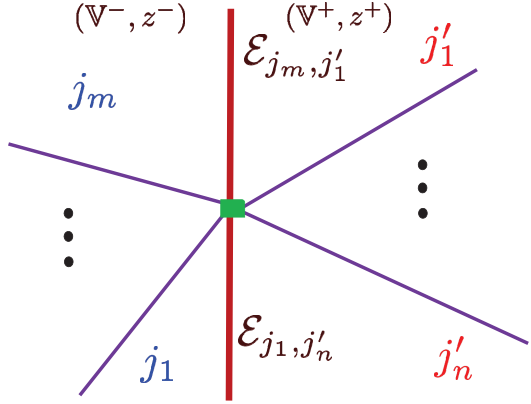}
\caption{\small Conventions for Chan-Paton factors localized on  interfaces. If representation spaces are attached
to the rays then this figure would represent a typical summand in
$\Hom(j_m j_1', j_1 j_n')$. We order such vertices from left to right using the conventions
of positive half-plane webs.  }
\label{fig:DOMAINWALL-CHANPATON}
\end{figure}

Now the categorified spectrum generator is given by the product
\be\label{eq:Inf-Vac-Homs2}
\widehat{R}(\CE) = \left( \oplus_{i,i'} \CE_{ii'} ~ e_{ii}\otimes e_{i'i'} \right)
\left(  \SpecGen(\CT^-,\CH^-)^{\rm tr}  \otimes 1\right)\left( 1\otimes \SpecGen(\CT^+,\CH^+) \right)
\left( \oplus_{j,j'} \CE_{jj'}  ~ e_{jj}\otimes e_{j'j'} \right)^*
\ee
See Figure \ref{fig:DOMAINWALL-CHANPATON} for a typical summand.

Now an interface amplitude is a degree one element $\CB^{-,+}\in \widehat{R}(\CE)$
satisfying the \afty-MC equation:
\be
\rho(\ft^{-,+})\left( \frac{1}{1-\CB^{-,+}}; e^{\beta_-}; e^{\beta_+} \right) = 0
\ee
We define an \emph{Interface} to be a pair
\be
\fI^{-,+} = (\CE^{-,+}, \CB^{-,+})
\ee
and we can define an \afty-category of Interfaces, denoted
\be
\fB\fr(\CT^-, \CT^+).
\ee
The objects of $\fB\fr(\CT^-, \CT^+)$ are Interfaces, for \underline{some} choice of CP data and
 the space of morphisms between $\fI_2^{-,+} $ and
$\fI_1^{-,+}  $ is the natural generalization of
\eqref{eq:BHOM}:
\be
\Hop(\fI_1^{-,+} , \fI_2^{-,+}) := \left( \oplus_{i,i'} \CE^1_{ii'} ~ e_{ii}\otimes e_{i'i'} \right)
\left(  \SpecGen(\CT^-,\CH^-)^{\rm tr}   \otimes 1\right) \left( 1\otimes \SpecGen(\CT^+,\CH^+)\right)
\left( \oplus_{j,j'} \CE^2_{jj'}  ~ e_{jj}\otimes e_{j'j'} \right)^*.
\ee
The \afty-multiplications are given by the natural generalization of  equation \eqref{eq:BraneMultiplications}:
we just contract with the taut element $\ft_{\CH} \to \ft^{-,+}$ and saturate all interior vertices
with the left or right interior amplitude $\beta^-, \beta^+$.

\begin{figure}[h]
\centering
\includegraphics[scale=0.2,angle=0,trim=0 0 0 0]{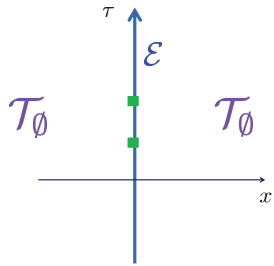}
\caption{\small The only taut interface web when $\CT^\ell, \CT^r$ are the trivial theory
has two boundary vertices. The boundary amplitude  is associated to a single
boundary vertex: $\CB\in \CE\otimes\CE^*  $  is a morphism of $\CE$ of degree
one. There is only one taut web, shown above. The MC therefore says that $\CB^2=0$.
Thus an  Interface between
the trivial theory and itself is the same thing as a chain complex.  }
\label{fig:INTFCE-TRIV-TRIV}
\end{figure}

\textbf{Remarks}:

\begin{enumerate}

\item An Interface between the empty theory and itself is precisely the data of a chain complex.
See Figure \ref{fig:INTFCE-TRIV-TRIV} for the explanation.

\item \emph{The identity Interface}.  A very useful example of an Interface is the
identity Interface $\fId \in \fB\fr(\CT,\CT)$. The CP spaces are
$\CE(\fId)_{ij} = \delta_{i,j} \IZ$  and
\be
\widehat{R}(\CE) = \oplus_{i,j } \widehat{R}^+_{ij} \otimes \widehat{R}^-_{ji} e_{ij}\otimes e_{ij}
\ee
where the superscripts $\pm$ indicate that $\widehat{R}$ is defined with respect to the positive, negative
half-plane, respectively.
To define the interface we take $\CB_{\CI}$ to have nonzero component only in summands of the form
$R_{ij}\otimes R_{ji}$ corresponding to the fan $\{i,j;j,i\}$.
The vertex looks like a straight line
of a fixed slope running through the domain wall. The boundary amplitude
 is  the element in $R_{ij}\otimes R_{ji}$   given by   $K_{ij}^{-1}$. and the Maurer-Cartan equation is proved
by Figure \ref{fig:ID-INTERFACE}:

\item  \emph{Landau-Ginzburg interfaces and branes in the product theory}: In the
context of Landau-Ginzburg models we can consider interfaces between a theory
defined by $(X_1,W_1)$ on the negative half-plane and $(X_2,W_2)$ on the positive
half-plane. By the doubling trick we would expect such interfaces to be related
to branes for the positive half-plane of the theory based on $(\bar X_1 \times X_2, \bar W_1 + W_2)$.
This is morally correct, but there are two closely related subtleties which should be pointed out.
First, from the purely abstract formalism, if we try to relate Interface amplitudes
for a pair of Theories $\CT^-, \CT^+$ to boundary amplitudes for $\CT^- \times \CT^+$
we will, in general,  fail: The vacua of the product theory are labeled by $(j_-, j_+)$ but the
slopes of the edges of the webs are the slopes of $z_{j_-^1, j_-^2} + z_{j_+^1, j_+^2}$.
In general half-plane fans for the product theory will have nothing to do with pairs
of half-plane fans in the left and right theories. The two concepts \underline{will} be equivalent,
however, in the special case that the web representations are of the form
\be\label{eq:SpecialRep}
R_{(j_-^1, j_+^1), (j_-^2, j_+^2)} = \delta_{j_-^1, j_-^2} R^+_{j_+^1,j_+^2} \oplus
\delta_{j_+^1, j_+^2} R^-_{j_-^1,j_-^2}.
\ee
Second, on the Landau-Ginzburg side, if we literally take the product metric and the
product superpotential then the Morse function $h_1 + h_2$ is too degenerate:
The critical manifolds are $\IR\times \IR$, corresponding to a center of mass
collective coordinate for two separate solitons. We must perturb the theory
by perturbing the superpotential with $\Delta W(\bar \phi_1, \phi_2)$.
Generic perturbations will in fact produce MSW complexes giving web representations
of the form \eqref{eq:SpecialRep}.

\end{enumerate}

\begin{figure}[h]
\centering
\includegraphics[scale=0.35,angle=0,trim=0 0 0 0]{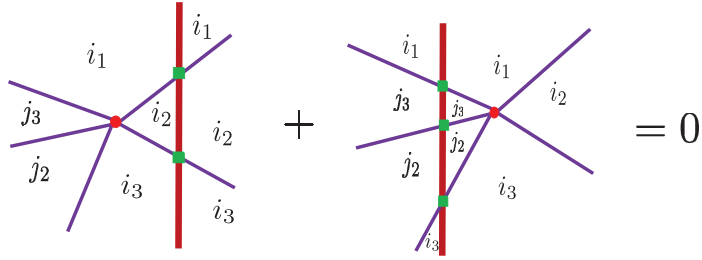}
\caption{\small Examples of taut interface webs which contribute to the
Maurer-Cartan equation for the identity interface $\fId$ between
a Theory and itself.    }
\label{fig:ID-INTERFACE}
\end{figure}

\subsubsection{Composition Of Interfaces}

\begin{figure}[h]
\centering
\includegraphics[scale=0.35,angle=0,trim=0 0 0 0]{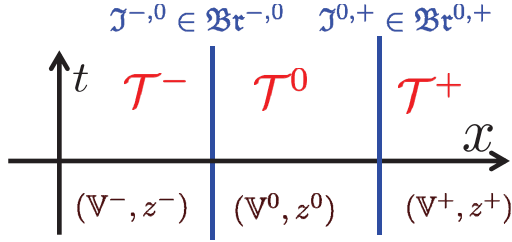}
\caption{\small Two Interfaces between a sequence of three Theories.   }
\label{fig:CompInterface1}
\end{figure}

\begin{figure}[h]
\centering
\includegraphics[scale=0.35,angle=0,trim=0 0 0 0]{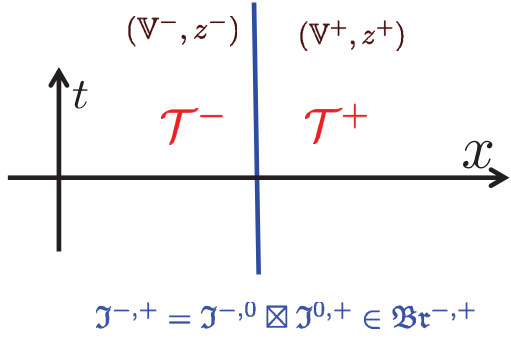}
\caption{\small  The Interface resulting from the ``operator product'' of the two
Interfaces.   }
\label{fig:CompInterface2}
\end{figure}

A crucial new ingredient is that Interfaces can be composed.
Suppose we have the situation shown in Figure \ref{fig:CompInterface1}
with a pair of Interfaces $\fI^{-,0}$ and $\fI^{0,+}$.
We will produce a new Interface,
denoted
\be
\fI^{-,0}\boxtimes \fI^{0,+}\in \fB\fr(\CT^-,\CT^+)
\ee
as shown in Figure \ref{fig:CompInterface2}.

\begin{figure}[h]
\centering
\includegraphics[scale=0.35,angle=0,trim=0 0 0 0]{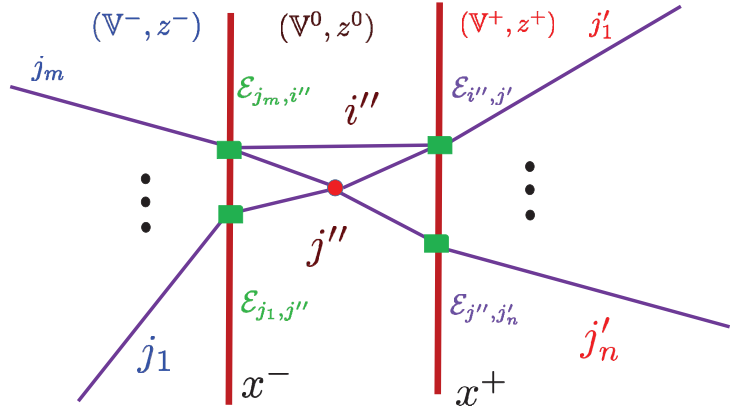}
\caption{\small An example of a composite web, together with conventions for
Chan-Paton factors. In this web the fan of vacua at
infinity has
$J_{\infty}(\fc) =  \{ j'_1, \dots j'_n; j_1, \dots, j_m \}$
Reading from left to right the indices are in clockwise order.   }
\label{fig:COMPOSITEWEB1}
\end{figure}

The key idea in the construction is to use ``composite webs''  $\fc = (\fu^-, \fs, \fu^+)$.
An example is shown in Figure \ref{fig:COMPOSITEWEB1}.
Again one can develop the whole web theory, write taut elements and a convolution
identity. (The convolution identity has some novel features. See \cite{Gaiotto:2015aoa}
for details.) The upshot is that the product Interface $\fI^{-,0}\boxtimes \fI^{0,+}$ has

\begin{enumerate}

\item \emph{Chan-Paton data}:

\be\label{eq:Comb-CP}
\CE(\fI^{-,0}\boxtimes \fI^{0,+})_{ii'}  :=
\oplus_{i'' \in \IV^0} \CE_{i, i''}^{-,0}\otimes \CE_{i'', i'}^{0,+}
\ee

\item \emph{Interface amplitude}:

\begin{equation}\label{eq:InterfaceComp}
\CB(\fI^{-,0}\boxtimes \fI^{0,+})   := \rho_\beta(\ft_c)\left[ \frac{1}{1-\CB^{-,0}}; \frac{1}{1-\CB^{0,+}}\right]
\end{equation}
where $\ft_c$ is the taut element for composite webs.

Using the convolution identity (omitted here) one can show that it indeed
satisfies the Maurer Cartan equations for an interface amplitude between the theories $\CT^-$ and $\CT^+$
with Chan-Paton spaces \eqref{eq:Comb-CP}.

\end{enumerate}

\begin{figure}[h]
\centering
\includegraphics[scale=0.49,angle=0,trim=0 0 0 0]{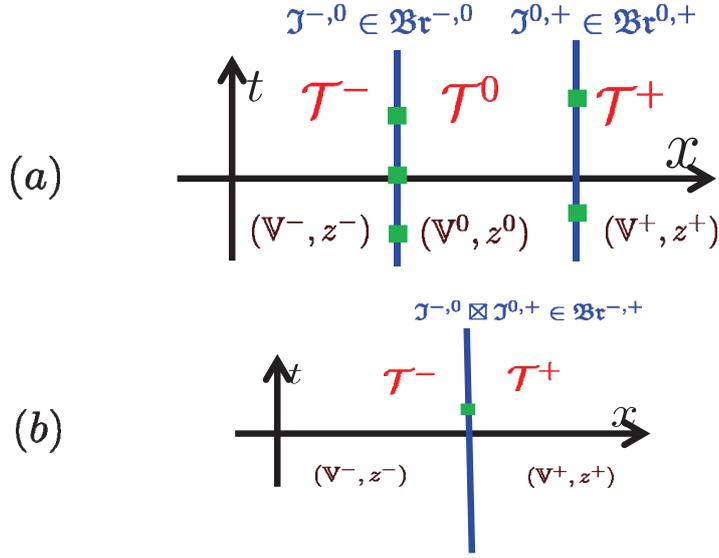}
\caption{\small Illustrating the bi-functor property: We take the ``OPE'' of both local boundary operators
on the interfaces, and of the interfaces, shown in (a),  to produce a local operator on an interface, shown in (b).   }
\label{fig:InterfaceBiFunctor}
\end{figure}

Now one can show that we have an \afty-bifunctor
\be
\fB\fr(\CT^-,\CT^0) \times \fB\fr(\CT^0,\CT^+) \to \fB\fr(\CT^-,\CT^+)
\ee
 This is illustrated in Figure \ref{fig:InterfaceBiFunctor}

An important
special case is that where $\CT^-$ is the trivial Theory so that
$\fB\fr(\CT^-,\CT^0)= \fB\fr(\CT^0)$. Then we see that a $\boxtimes$ with a fixed Interface
$\fI \in \fB\fr(\CT^0,\CT^+)$ gives an \afty-functor on categories of Branes:
\be
\fB\fr( \CT^0) \times \fB\fr(\CT^0,\CT^+) \to \fB\fr( \CT^+)
\ee
Physically, we are moving a $0,+$ interface $\fI$ into a boundary and mapping
a boundary condition for Theory $\CT^0$ to one for Theory $\CT^+$.

Thus, our quest for parallel transport of Brane categories will be fulfilled
if we can find suitable Interfaces $\fI[\wp]$ associated with paths between
theories $\CT^{\ell}$ and $\CT^{r}$.

\subsubsection{An $A_\infty$ 2-Category Of Interfaces }

A natural question to ask about the composition of Interfaces is whether it is associative.
In fact, to define the composite webs we need to choose positions on the $x$-axis for
the two domain walls as well as the position of the final interface. These positions
can influence the set of composite webs.  So we should really
denote the product of Interfaces by
\be
\left( \fI^{-,0} \IntfcTimes \fI^{0,+}\right)_{x^{-,0}, x^{0,+}, x^{-,+} }
\ee
However, one can show that the product only depends on these positions up to ``homotopy equivalence.''
The proof, which is somewhat long, involves developing a theory of webs which are time-dependent.
Similarly, one can prove that the composition is associative, up to ``homotopy equivalence.''
All the details are in \cite{Gaiotto:2015aoa}.

To define ``homotopy equivalence'' let us note that the
 \afty-structure of the category of Branes and Interfaces requires in part that
the Hop spaces have a differential: If $\delta\in \Hop(\fB_1, \fB_2)$ then
\be
M_1(\delta) = \rho_{\beta}(\ft_{\CH}) \left( \frac{1}{1-\CB_1}, \delta, \frac{1}{1-\CB_2} \right)
\ee
and $M_1\circ M_1 =0$, when the Hop spaces are composable. We can thus define a notion of homotopy equivalence
of Branes (and entirely parallel definitions apply to Interfaces):

1. Two morphisms are homotopy equivalent if $\delta_1 - \delta_2 = M_1(\delta_3) $.

2. Two Branes  are homotopy equivalent, denoted,
 $\CB \sim \CB', $ if there are two $M_1$-closed morphisms $\delta: \CB \to \CB'$ and $\delta': \CB' \to \CB$ which
are inverses up to homotopy. That is:
\begin{equation}\label{eq:hmtpy-Br}
M_2(\delta, \delta') \sim \Id \qquad \qquad M_2(\delta', \delta) \sim \Id.
\end{equation}
where $\Id$ is the natural identity in $\oplus_i \CE_i \otimes \CE_i^*$.

The net result of these observations
 is that we have defined what might be called an ``\afty-2-category'' structure:

\begin{enumerate}

\item The objects, or $0$-morphisms are the Theories.

\item The $1$-morphisms between two Theories are Interfaces $\fI^{-,+}$.

\item The $2$-morphisms between two $1$-morphisms are the boundary-changing
operators on the Interface.

\end{enumerate}

\begin{figure}[h]
\centering
\includegraphics[scale=0.35,angle=0,trim=0 0 0 0]{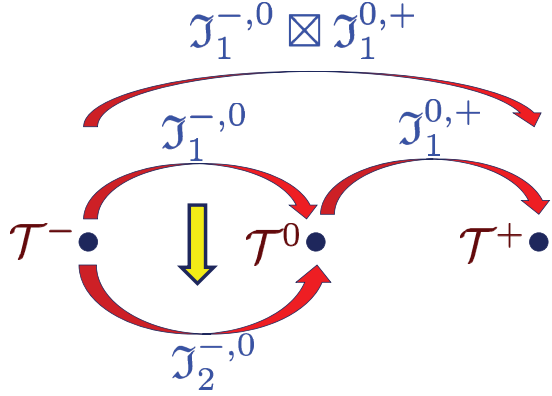}
\caption{\small Illustrating the two category of Theories, Interfaces, and boundary operators.    }
\label{fig:TwoCategoryInterfaces}
\end{figure}

This is illustrated in Figure \ref{fig:TwoCategoryInterfaces}.

\subsection{An Example Of Categorical Transport}

We will now sketch how one can actually construct a parallel transport interface
for a tame vacuum homotopy:
\be\label{eq:SpinningWeights}
\wp: x\mapsto \{ z_i(x)\} \in \IC^N - \fE
\ee
which does not cross the exceptional walls $\fE$.  We assume $\wp(x)$ only varies on a
compact set $[x_{\ell}, x_r]$.

Our goal is to define an Interface
\be
\fI[\wp] \in \fB\fr(\CT^{\ell}, \CT^{r})
\ee
so that if $\wp^1(x) \sim \wp^2(x)$ give homotopic paths of vacuum weights with fixed
endpoints then $\fI[\wp^1]$ and $\fI[\wp^2]$ are homotopy-equivalent
Interfaces, and such that if we compose two paths then
\be\label{eq:trspt-1}
\fI[\wp^1]\IntfcTimes \fI[\wp^2] \sim \fI[\wp^1\circ \wp^2]
\ee
where $\sim$ means homotopy equivalence.

\begin{figure}[h]
\centering
\includegraphics[scale=0.20,angle=0,trim=0 0 0 0]{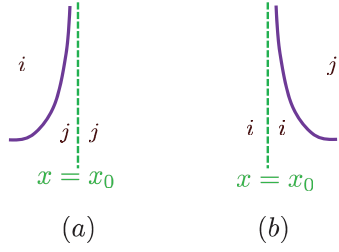}
\caption{\small Near a future stable binding point $x_0$ of type $ij$ the edges separating
 vacuum $i$ from $j$ asymptote to the dashed green line $x=x_0$ in the future. Figures (a) and (b)  show
 two possible behaviors of such lines. The phase $e^{-\I \vartheta(x)} z_{ij}$ rotates through the
 positive imaginary axis in the counterclockwise direction.    }
\label{fig:FUTURESTABLE-1}
\end{figure}

\begin{figure}[h]
\centering
\includegraphics[scale=0.20,angle=0,trim=0 0 0 0]{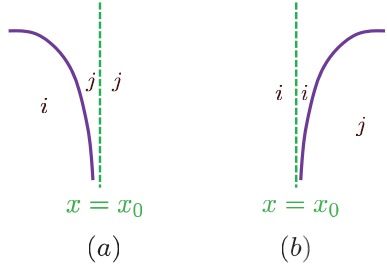}
\caption{\small Near a past stable binding point $x_0$ of type $ij$ the edges separating
 vacuum $i$ from $j$ asymptote to the dashed green line $x=x_0$ in the past. Figures (a) and (b)  show
 two possible behaviors of such lines. The phase $e^{-\I \vartheta(x)} z_{ij}$ rotates through the
 positive imaginary axis in the clockwise direction.   }
\label{fig:PASTSTABLE-1}
\end{figure}

The key is to construct an analogous theory of \emph{curved webs} where the $ij$ edges
have tangents at $(x,\tau)$ parallel to $z_i(x) - z_j(x)$. One crucial new feature
emerges for curved webs. Following the tangent vectors, sometimes the edges are
forced to go to infinity at finite values of $x$. These special values of $x$ are known
as \emph{binding points}. We can have ``future stable'' binding points as in
Figure \ref{fig:FUTURESTABLE-1} or  ``past stable'' binding points as in
Figure \ref{fig:PASTSTABLE-1}.


The binding points $x_0$ are characterized as the values of $x$ for which
\be\label{eq:Sij-Ray-def}
z_{ij}(x_0) \in \I \IR_+
\ee
The future/past stability is determined by the sense in which $\Re\left(   z_{ij}(x) \right)$
passes through zero as $x$ passes through $x_0$:

\begin{enumerate}

\item \emph{Future stable binding point}: As $x$ \underline{increases} past $x_0$
 $z_{ij}(x)$ goes through the positive imaginary axis in the \underline{counter-clockwise}
 direction.

\item \emph{Past stable binding point}: As $x$ \underline{increases} past $x_0$
 $z_{ij}(x)$ goes through the positive imaginary axis in the \underline{clockwise}
 direction.

\end{enumerate}

Now we define Chan-Paton data of the desired Interface.
For each  binding point $x_0$ of type $ij$   introduce a matrix
with chain-complex entries. It depends on whether $x_0$ is future-stable or past-stable:
\be\label{eq:SijFactor-def-fs}
S_{ij}(x_0) := \IZ\cdot \textbf{1} + R_{ij} e_{ij}\qquad \qquad future\  stable
\ee
\be\label{eq:SijFactor-def-ps}
S_{ij}(x_0) := \IZ\cdot \textbf{1} + R_{ji}^* e_{ij}\qquad \qquad past\  stable.
\ee
We will refer to $S_{ij}(x_0)$  as a \emph{categorified $S_{ij}$-factor},
or just as an \emph{  $S_{ij}$-factor}, for short. Then we define the
Chan-Paton factors of the Interface to be:
\be\label{eq:TautCurvedCP}
 \oplus_{i,j\in \IV} \CE_{i,j} e_{i,j} := \bigotimes_{i\not=j}
\bigotimes_{x_0 \in \curlyvee_{ij} ~ \cup \curlywedge_{ij} } S_{ij}(x_0)
\ee
where the tensor product on the RHS of \eqref{eq:TautCurvedCP} is an ordered product
over binding points, ordered from left to right by
increasing values of $x_0$. The amplitudes for the Interface are simply given by evaluating
the taut curved web on the interior amplitude:
$\rho(\ft_{\rm curved})(e^{\beta})$. (This formula needs some interpretation. See
\cite{Gaiotto:2015aoa} for details.) In this way we get an Interface
\be\label{eq:defIth}
 \fI[\wp] \in \fB\fr(\CT^{\ell}, \CT^{r})
\ee
associated to the tame vacuum homotopy $\wp(x)$.  It satisfies the desired properties for
flat parallel transport.

\begin{figure}[h]
\centering
\includegraphics[scale=0.28,angle=0,trim=0 0 0 0]{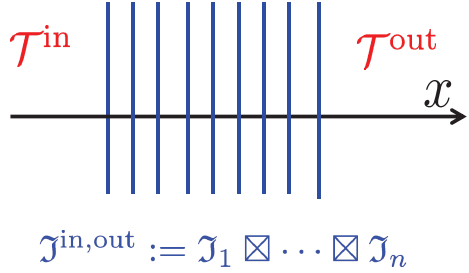}
\caption{\small Breaking up the path $\wp$ into elementary paths we need only
produce special interfaces for ``trivial'' transport, and for transport
across $S$-walls.   }
\label{fig:ElementaryInterfaces}
\end{figure}

Note that,  thanks to the composition property \eqref{eq:trspt-1}, up to homotopy we can
break up $ \fI[\wp]$ as a product of Interfaces as in Figure \ref{fig:ElementaryInterfaces}.
Therefore to construct $\fI[\wp]$ we need only construct then the Interfaces for crossing the $S_{ij}$ walls.
These are denoted $\fS^{p,f}_{ij}$ for past and future stable crossings,
respectively. The amplitudes can be described quite explicitly  \cite{Gaiotto:2015aoa}.
The functors $\fB \to \fB \IntfcTimes \fS^{p,f}_{ij}$ are closely
related to mutations.

\subsubsection{Categorified S-Wall-Crossing}\label{subsubsec:Cat-S-Wall}

We now return to one of our motivations from Section \S \ref{subsec:Goals} above,
namely the categorification of the $S$-wall crossing that plays such an important
role in the theory of spectral networks \cite{Gaiotto:2012rg}.
Given an Interface $\fI^{-,+}$ associated with a path of theories
the \emph{framed BPS degeneracies} are, by definition:
\be\label{eq:FramedBPS-def}
\fro(\fI^{-,+}, ij'):= \Tr_{\CE( \fI^{-,+})_{ij'}} e^{\I \pi {\rm \textbf{F}}}.
\ee
If we consider a path $\wp_x$ whose endpoint terminates with $z(x)$, and that crosses an
 $ij$ binding point as $x$ increases past $x_0$  (and hence $z(x)$ crosses
an $S_{ij}$-wall)   then the
matrix of Witten indices
\be
F[\wp_x] := \sum_{k,\ell} \fro(\fI[\wp_x], k,\ell) e_{k,\ell}.
\ee
jumps by
\be
F \mapsto \begin{cases} F \cdot (\textbf{1} + \mu_{ij} e_{ij} ) &   x_{ij} \in \curlywedge_{ij} \\
F \cdot (\textbf{1} - \mu_{ji} e_{ij} ) &   x_{ij} \in \curlyvee_{ij} \\
\end{cases}
\ee
according to whether the binding point is future or past stable, respectively.
This is the framed wall-crossing formula. Now, since the Witten index of $R_{ij}$ is $\mu_{ij}$ we recognize the
formula for the change of the Interface
\be
\fI[\wp_x] \to  \fI[\wp_x] \boxtimes \fS^{p,f}_{ij}
\ee
as $x$ crosses the binding point as a categorification of the $S$-wall crossing formula.

\bigskip
\noindent
\textbf{Example}: Consider a theory with two vacua, such as the Landau-Ginzburg model with
$W\sim  \phi^3 - z \phi$. The family is parametrized by $z\in C$ with $C=\IC^*$.  There are two massive vacua
at $\phi_{\pm } = \pm z^{1/2}$.  We choose a path $\wp$ defined by
$z(x)$ in $\IC^*$ where $x\in [\epsilon, 1-\epsilon]$
for $\epsilon$ infinitesimally small and positive with $z(x) = e^{\I (1-2x) \pi}$. Thus the path nearly
encircles the singular point $z=0$ beginning just above and ending just below the cut for the principal
branch of the logarithm. If $\zeta$ has a small positive phase then there   are two binding points of type $+-$
at $x= 1/3-\delta , 1-\delta$ and one binding point of type $-+$ at $x=2/3-\delta$ where we can take $\delta$ samll
with $\delta > \epsilon$.
These binding points are all
future stable. The wall-crossing formula for the framed BPS indices amounts to a simple matrix identity:
\be\label{eq:Spl-Fr-WC}
\begin{pmatrix} 1 & 0 \\  1 & 1 \\ \end{pmatrix}
\begin{pmatrix} 1 & -1 \\  0 & 1 \\ \end{pmatrix}
\begin{pmatrix} 1 & 0 \\  1 & 1 \\ \end{pmatrix}
=
\begin{pmatrix} 0 & -1 \\  1 & 0 \\ \end{pmatrix}
\ee
where the three factors on the LHS reflect the wall-crossing across the three $S_{ij}$-rays,
and the matrix on the RHS accounts for the monodromy of the vacua.
The categorification of the wall-crossing identity \eqref{eq:Spl-Fr-WC},
at least at the level of Chan-Paton factors,  is obtained by replacing the
matrix of Witten indices on the LHS of \eqref{eq:Spl-Fr-WC} by the Chan-Paton
factors of the three Interfaces of type $\fS$ to get:
\be\label{eq:CP-prod}
\begin{split}
\begin{pmatrix}\IZ & 0 \\  \IZ[f_2] ~~ & \IZ \\ \end{pmatrix}
\begin{pmatrix} \IZ & ~~ \IZ[f_1] \\  0 & \IZ \\ \end{pmatrix}
\begin{pmatrix}\IZ & 0 \\  \IZ[f_2]~~ & \IZ \\ \end{pmatrix}
& =
\begin{pmatrix}\CE_{--}   & \CE_{-+} \\
\CE_{+-}  & \CE_{++}  \\ \end{pmatrix}\\
\end{split}
\ee
where $f_1, f_2$ are integral fermion number shifts and $f_1 + f_2=1$.
Multiplying out the matrices we see that  $\CE_{-+}= \IZ[f_1]$, while
\be
\CE_{--} = \CE_{++}= \IZ \oplus \IZ[1]
\ee
is a complex with a degree one differential (we have used $f_1+f_2 =1$) and
\be
\CE_{+-} =  \IZ[f_2]\oplus \IZ[f_2]\oplus \IZ[f_2+1]
\ee
is another complex with a degree one differential.  The matrix of complexes \eqref{eq:CP-prod} is
quasi-isomorphic to the categorified version of the monodromy:
\be
\begin{pmatrix} 0 & \IZ[ 1-f_2 ] \\  \IZ[f_2] & 0 \\ \end{pmatrix}.
\ee
\begin{figure}[h]
\centering
\includegraphics[scale=0.25,angle=0,trim=0 0 0 0]{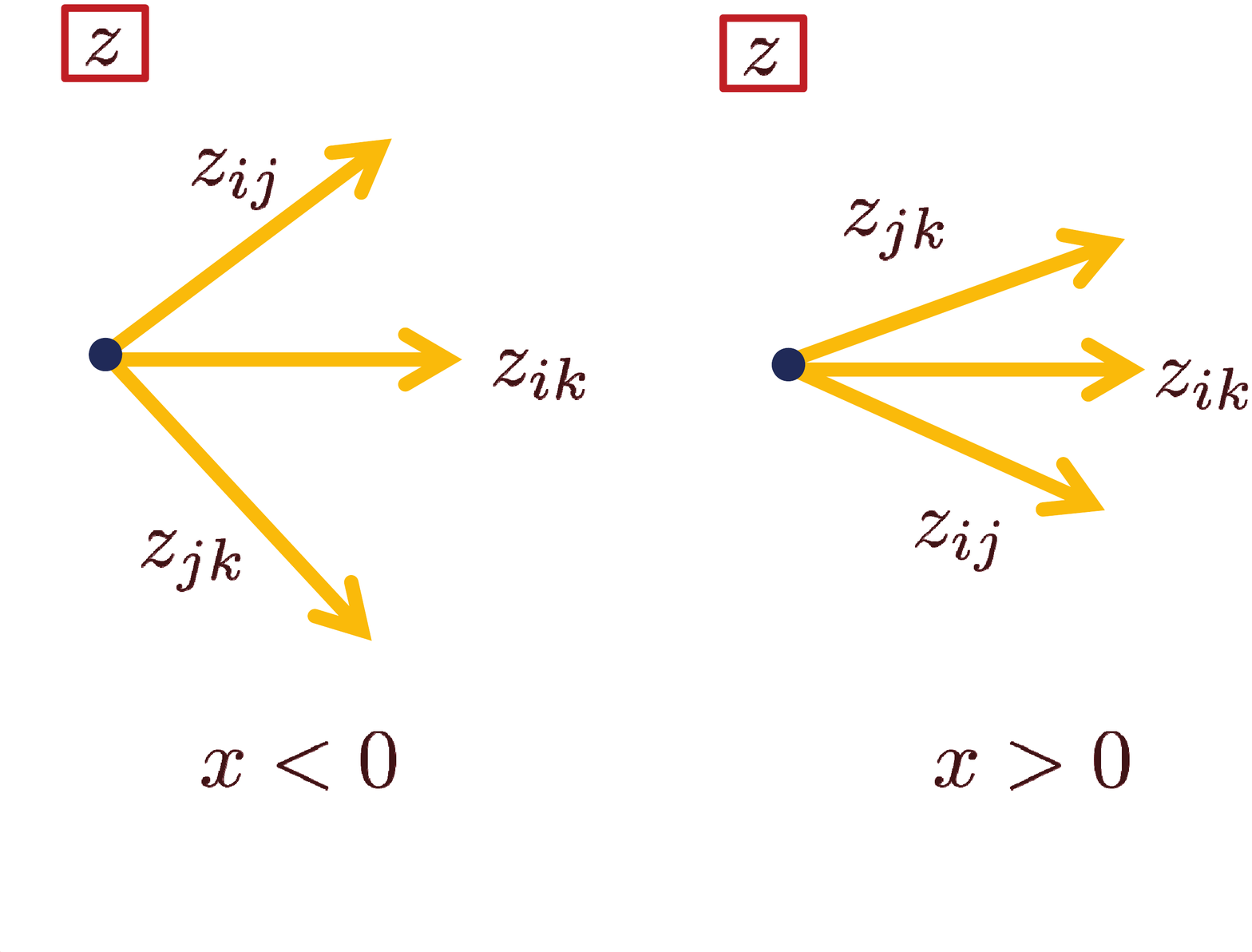}
\caption{\small For the path of vacuum weights in Figure
%
\protect\ref{fig:CAT-CVWC-1}
we have BPS rays crossing as in the standard marginal stability
analysis of the two-dimensional wall-crossing formula.   }
\label{fig:CAT-CVWC-BPSRAYS}
\end{figure}
\begin{figure}[h]
\centering
\includegraphics[scale=0.35,angle=0,trim=0 0 0 0]{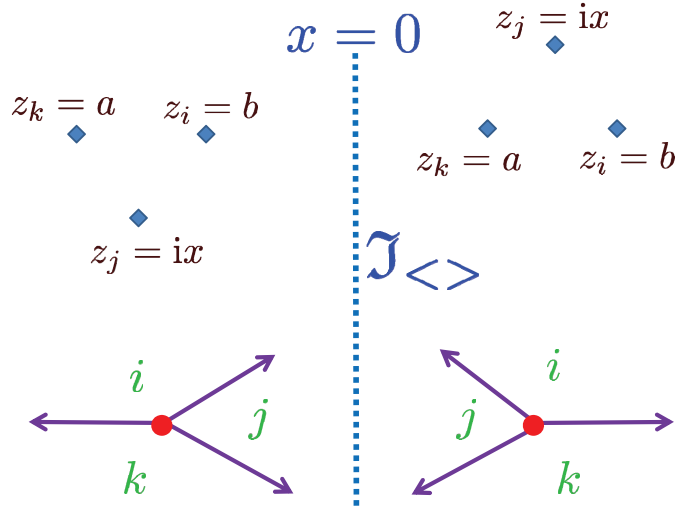}
\caption{\small An example of a continuous path of vacuum weights crossing a wall of
marginal stability. Here $z_k=a$ and $z_i=b$ with $a,b$ real and $a<0<b$.
They do not depend on $x$, while $z_j(x) = \I x$. We show typical vacuum weights
for negative and positive $x$ and the associated trivalent vertex.   As $x$ passes
through zero the vertex  degenerates with $z_{jk}(x)$  and $z_{ij}(x)$ becoming real.
Note that with this path of weights the $\{i,j,k\}$ form a \emph{positive} half-plane
fan in the negative half-plane, while $\{k,j,i\}$ form a \emph{negative} half-plane
fan in the positive half-plane. If we choose $x_\ell < 0 < x_r$ there is an
associated interface $\fI_{<>}$.    }
\label{fig:CAT-CVWC-1}
\end{figure}
\begin{figure}[h]
\centering
\includegraphics[scale=0.35,angle=0,trim=0 0 0 0]{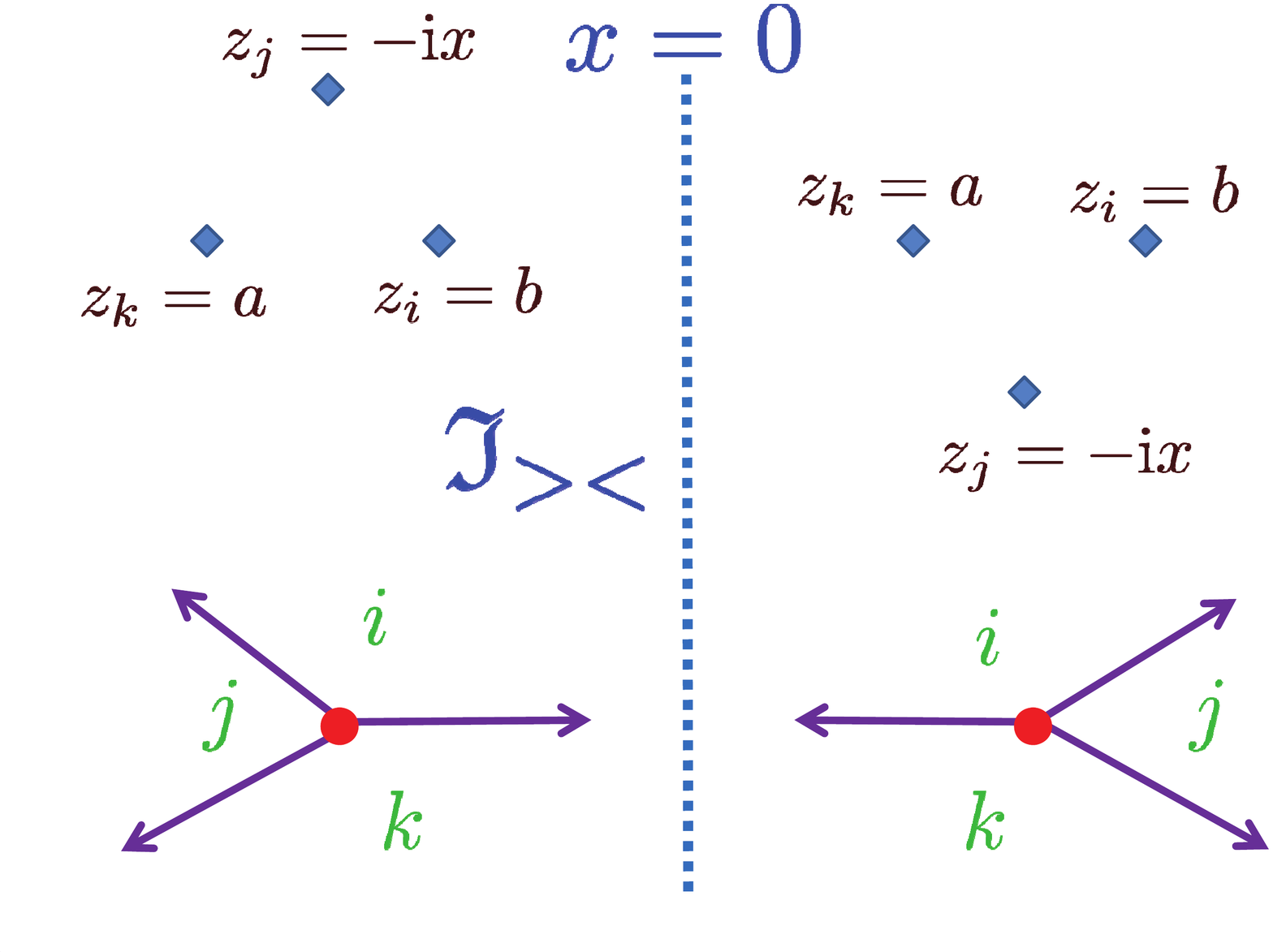}
\caption{\small In this figure the path of weights shown in Figure
%
%
\protect\ref{fig:CAT-CVWC-1}
 is reversed.
Again,  $z_k=a$ and $z_i=b$ with $a,b$ real and $a<0<b$, but now   $z_j(x) = - \I x$.
We show typical vacuum weights
for negative and positive $x$ and the associated trivalent vertex.
Note that with this path of weights the $\{i,j,k\}$ form a \emph{positive} half-plane
fan in the positive half-plane, while $\{k,j,i\}$ form a \emph{negative} half-plane
fan in the negative half-plane. In order to define an interface we choose initial
and final points for the path $-x_r < 0 < - x_\ell$ so that, after translation,
it can be composed with the path defining $\fI_{<>}$.
%
%
%
   }
\label{fig:CAT-CVWC-2}
\end{figure}

\subsection{Categorified Wall-Crossing For 2d Solitons}\label{subsec:Cat-CVWC}

The standard wall-crossing formula for BPS indices of 2d solitons was
studied by Cecotti and Vafa in \cite{Cecotti:1992qh}. It is associated with a homotopy
of vacuum weights so that the cyclic orders of the central charges gets
reversed, as in Figure \ref{fig:CAT-CVWC-BPSRAYS}.
We can realize this by the explicit homotopy of vacuum weights shown in
Figures \ref{fig:CAT-CVWC-1} and \ref{fig:CAT-CVWC-2}.
The wall-crossing of the BPS indices is a special case of the famous Kontsevich-Soibelman
wall-crossing formula:

\be\label{eq:CV-KS-WC}
(1+\mu_{ij}^{(1)} e_{ij}) (1+\mu_{ik}^{(1)} e_{ik}) (1+\mu_{jk}^{(1)} e_{jk}) =
(1+\mu_{jk}^{(2)} e_{jk}) (1+\mu_{ik}^{(2)} e_{ik}) (1+\mu_{ij}^{(2)} e_{ij})
\ee
which gives:
\be\label{eq:W-indx-wc}
\begin{split}
\mu_{ij}^{(2)}  &  = \mu_{ij}^{(1)} \\
  \mu_{jk}^{(2)}  &   = \mu_{jk}^{(1)} \\
\mu_{ik}^{(2)}  & = \mu_{ik}^{(1)} + \mu_{ij}^{(1)} \mu_{jk}^{(1)} . \\
\end{split}
\ee

To categorify this we seek to define Interfaces:
\be
\fI_{<>} \in \fB\fr(\CT^\ell, \CT^r) \qquad\qquad \&  \qquad\qquad \fI_{><} \in \fB\fr(\CT^r, \CT^\ell)
\ee
(where the  notation is meant to remind us how the half-plane fans are configured in
the negative and positive half-planes).
Now, the essential statement constraining these Interfaces is that  the
composition of the Interfaces should be homotopy equivalent to the identity Interface:
\be\label{eq:Cat-WC-Form1}
\fI_{<>}\IntfcTimes \fI_{><} \sim \fId_{\CT^\ell} \qquad \qquad \&  \qquad \qquad  \fI_{><}\IntfcTimes \fI_{<>} \sim \fId_{\CT^r}.
\ee

In \cite{Gaiotto:2015aoa} we  construct such Interfaces $\fI_{><}$ and $\fI_{<>}$
and show that the most natural solution to the constraints follows from:
\be\label{eq:Cat-2dwc}
\begin{split}
R_{ij}^{(2)}   &  = R_{ij}^{(1)} \\
 R_{jk}^{(2)}   &  = R_{jk}^{(1)} \\
R^{(2)}_{ik} - R^{(1)}_{ik} & =  \left(R_{ij}\otimes R_{jk}\right)^+ - \left(R_{ij}\otimes R_{jk}\right)^-\\
& = \left( R_{ij}^+ - R_{ij}^-\right)\otimes \left( R_{jk}^+ - R_{jk}^- \right) \\
\end{split}
\ee
where the superscript $\pm$ on the right hand side refers to the sign of $(-1)^{{\rm \textbf{F}}}$.
We have written an identity of virtual vector spaces. One could move terms to left and
right hand sides so that only plus signs appear and we would then have an identity of vector spaces.
We have written the equation in terms of virtual vector spaces to bring out the fact that
\eqref{eq:Cat-2dwc} is a categorification of the wall-crossing formulae
\eqref{eq:W-indx-wc}.

\begin{figure}[h]
\centering
\includegraphics[scale=0.35,angle=0,trim=0 0 0 0]{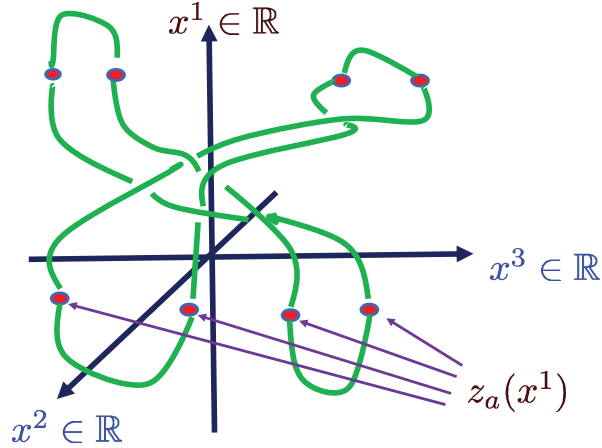}
\caption{\small This figure depicts the a knot (actually, a link) in the boundary at $y=0$ at a fixed value of $x^0$.
It is presented as a tangle evolving in the $x^1$ direction and therefore can be characterized
as a trajectory   of points $z_a(x^1)$ in the complex $z=x^2 + \I x^3$ plane.  The tangle is closed
by ``creation'' and ``annihilation'' of the points $z_a$  in pairs (with identical values of $k_a$).     }
\label{fig:KNOT-HOM-4}
\end{figure}

\subsection{Potential Application To Knot Homology}\label{subsec:KnotHomology}

To conclude, let us consider very briefly the motivation from knot homology.
For background see \cite{Witten:2011zz}\cite{WItten:2011pz}\cite{Witten:2014xwa}\cite{Gaiotto:2011nm},
and the review in \S 18.4  of \cite{Gaiotto:2015aoa}.  The central idea is to consider
five-dimensional supersymmetric gauge theory on a five-manifold with boundary:
\be
M_5 = \IR \times M_3\times \IR_+,
\ee
where $M_3$ is a three-manifold. The knot resides in $M_3$ on the boundary and is
used to formulate the
crucial boundary conditions for the instanton equations of the gauge theory. These 5d
instanton equations were first written in \cite{Haydys,Witten:2011zz}. At a formal
level they turn out to be the $\zeta$-instanton
equations for a gauged Landau-Ginzburg model whose target space is a space of complexified
gauge connections on $M_3$ \cite{Gaiotto:2015aoa}. In the case when $M_3 = \IR \times C$,
with $C$ a Riemann surface,
the equations are also the $\zeta$-instanton equations for a gauged Landau-Ginzburg model
whose target space is a space of complexified gauge fields on $\tilde M_3 = C \times \IR_+$.
It is this latter form which forms the background for the discussion of \cite{Gaiotto:2011nm}.
In either case, the knot complex is the MSW complex for the Landau-Ginzburg theory.

When $M_5 = \IR \times \IR \times C \times \IR_+$ we denote coordinates on the first two
factors by $(x^0,x^1)$. We consider the case where the
knot is in $\IR^3$ (so $C$ is just the complex plane)
and  is furthermore presented as a tangle, i.e.  an evolving
set of points in the complex plane,  $z_a(x^1)$, $a=1,\dots n$, as in Figure \ref{fig:KNOT-HOM-4}.

%

For any collection $S$ of strands parallel to the $x^1$ axis
\begin{itemize}
\item Solutions of the 5d instanton equation which do not depend on $(x^0,x^1)$ will give some vacuum data $\IV_S$.
\item Solutions of the 5d instanton equation which depend only on the combination $x^1 \cos \mu + x^0 \sin \mu$ will provide the spaces of
solitons which can interpolate between any two given vacua and thus web representations for the vacuum data $\IV_S$.
\item Solutions of the 5d instanton equation with fan-like asymptotics in the $(x^0,x^1)$ plane will provide interior amplitudes $\beta_S$
and thus Theories $\CT_S$.
\end{itemize}
If $S$ is an empty collection, we expect the theory $\CT_S$ to be trivial.

Similarly, for any ``supersymmetric interface'' $\CI$, i.e. a time-independent boundary condition for the 5d equations which involves
a set of parallel strands $S^-$ for $x^1\ll -L$ and a set of parallel strands $S^+$ for $x^1\gg L$
\begin{itemize}
\item Solutions of the 5d instanton equation which do not depend on time will give Chan-Paton data $\CE^\CI_{j,j'}$.
\item Solutions of the 5d instanton equation with fan-like asymptotics in the $(x^0,x^1)$ plane will provide boundary amplitudes $\CB^\CI$
and thus an Interface $\fI[\CI]$ between Theories $\CT_{S^-}$ and $\CT_{S^+}$.
\end{itemize}

We can assume that the stretched link is approximated by a sequence of collections of strands $S_a$, starting and ending with the empty collection $S_0 = S_n = 0$,
separated by interfaces $\CI_{a,a+1}$. The approximate ground states and instantons of the knot homology complex
will literally coincide with the chain complex of the Interface $\fI({\rm Link})$ between the trivial Theory and
itself, defined as the composition of the Interfaces $\fI[\CI_{a,a+1}]$
\be\label{eq:Link-Interface0}
\fI({\rm Link}) :=  \fI[\CI_{0,1}]\boxtimes \cdots \boxtimes  \fI[\CI_{n-1,n}].
\ee

Furthermore, if we allow the transverse position of the strands to evolve adiabatically in between discrete events such as recombination of strands,
according to some profile $S_a(x^1)$, we expect the knot homology complex to coincide with the chain complex of an Interface $\fI({\rm Link})$
which include the insertion of the corresponding categorical parallel transport Interfaces:
\be\label{eq:Link-Interface0}
\fI({\rm Link}) :=  \fI[\CI_{0,1}]\boxtimes \fI[\CT_{S_1(x^1)}]\boxtimes \cdots \boxtimes\fI[\CT_{S_{n-1}(x^1)}] \boxtimes  \fI[\CI_{n-1,n}].
\ee
We conjecture that the chain complexes so constructed define a knot homology theory.
The required double-grading comes about as follows: The $R_{ij}$ and Chan-Paton
data have the usual grading by ${\rm \textbf{F}}$. The second grading comes from the fact that the
relevant superpotential $W$ is a Chern-Simons functional. In particular, it is not
single valued and $dW$ can have interesting periods.

\section*{Acknowledgements}

We would especially like to thank Nick Sheridan for many useful
discussions, especially concerning the Fukaya-Seidel category. We would also like to thank M. Abouzaid, K. Costello, T. Dimofte,
D. Galakhov, E. Getzler, M. Kapranov, L. Katzarkov, M. Kontsevich,  Kimyeong Lee,  S. Lukyanov, Y. Soibelman, and A. Zamolodchikov
for useful discussions and  correspondence.
The research of DG was supported by the
Perimeter Institute for Theoretical Physics. Research at Perimeter Institute is supported
by the Government of Canada through Industry Canada and by the Province of Ontario
through the Ministry of Economic Development and Innovation.
The work of   GM is supported by the DOE under grant
DOE-SC0010008 to Rutgers and NSF Focused Research Group award DMS-1160591.  GM also gratefully acknowledges
the hospitality of  the Institute for Advanced Study, the Perimeter Institute for Theoretical Physics,
the Aspen Center for Physics  (under
NSF Grant No. PHY-1066293), the
KITP at UCSB (NSF Grant No. NSF PHY11-25915) and the
Simons Center for Geometry and Physics. He also thanks the organizers of the
\emph{Journ\'ees de Physique Math\'ematique, Lyon, 2014}  and the
Conference on Homological Mirror Symmetry and Geometry in Miami for the
opporunity to present lecture series on this material.
The work of EW is supported in part
by NSF Grant PHY-1314311.

\end{document}